\documentclass[a4paper, 11pt]{article}

\usepackage{amsmath, amssymb, anysize, graphicx,
	latexsym, stmaryrd, xspace,amsfonts}
\usepackage{amsthm}
\usepackage[ngerman,english]{babel}    
\usepackage[right]{eurosym}
\usepackage{csquotes}
\usepackage{array}
\usepackage[normalem]{ulem}
\usepackage{cancel}
\usepackage{tikz}
\usetikzlibrary{positioning}
\usetikzlibrary{arrows}
\usetikzlibrary{shapes}

\usepackage{helvet}
\usepackage{color}
\usepackage{empheq}

\usepackage{fancyhdr, lastpage}
\usepackage{hyperref}
\usepackage{cases}
\usepackage{mathabx}
\usepackage{xfrac}

\usepackage{wrapfig}
\usepackage{figbib}
\usepackage{cite}

\usepackage{enumerate}
\usepackage[capitalise]{cleveref}
\usepackage{verbatim}

\usepackage{pgfplots}

\usetikzlibrary{positioning}
\usetikzlibrary{arrows}
\usetikzlibrary{shapes}
\pgfplotsset{compat=newest}
\usetikzlibrary{plotmarks}
\usetikzlibrary{arrows.meta}
\usetikzlibrary{matrix}
\usetikzlibrary{pgfplots.groupplots}
\pgfplotsset{plot coordinates/math parser=false}
\pgfplotsset{try min ticks=3}
\pgfplotsset{plot coordinates/math parser=false}
\newlength\figureheight
\newlength\figurewidth

\usepackage[a4paper, left=2.2cm, right=2.0cm, top=2.5cm,bottom=2cm]{geometry}
\newtheorem{theo}{Theorem}[section]
\newtheorem{lem}[theo]{Lemma}

\newtheorem{remark}[theo]{Remark}

\newcommand{\mysection}[1]{\section{#1} \setcounter{equation}{0}}

\newcounter{expcounter}			
\newenvironment{experiment}
{\refstepcounter{expcounter}
	{~\\\textbf{Experiment \theexpcounter~---~}}
}

\newcommand{\dd}{\,\mathrm{d}}

\newcommand{\R}{\mathbb{R}}

\newcommand{\eps}{\varepsilon}

\newcommand{\Om}{\Omega }

\newcommand{\cb}{}

\title{Multiscale modeling of glioma invasion: from receptor binding to flux-limited macroscopic PDEs}

\author{Anne Dietrich$^{1}$, \quad  Niklas Kolbe$^{2}$, \quad Nikolaos Sfakianakis$^{3}$, \quad Christina Surulescu$^{1}$,\\
	{\small $^{1}$ Technische Universit\"{a}t Kaiserslautern, Felix-Klein-Zentrum f\"{u}r Mathematik,} \\
	{\small Paul-Ehrlich-Str. 31, 67663 Kaiserslautern, Germany}\\
	{\small $^{2}$ Kanazawa University, Faculty of Mathematics \& Physics,}\\  
	{\small Kakuma, Kanazawa 920-1192, Japan}\\
	{\small $^{3}$ University of St. Andrews, School of Mathematics \& Statistics,}\\
	{\small North Haugh, St. Andrews, Fife, KY16 9SS, Scotland, UK}\\
				{\small (adietric@mathematik.uni-kl.de, kolbe@staff.kanazawa-u.ac.jp,}\\ {\small n.sfakianakis@st-andrews.ac.uk,   surulescu@mathematik.uni-kl.de)}
}
\begin{document}
	\date{}
	\maketitle
	
	
	\begin{abstract}
	We propose a novel approach to modeling cell migration in an anisotropic environment with biochemical heterogeneity and interspecies interactions, using as a paradigm glioma invasion in brain tissue under the influence of hypoxia-triggered angiogenesis. The multiscale procedure links single-cell and mesoscopic dynamics with  population level behavior, leading on the macroscopic scale to flux-limited glioma diffusion and multiple taxis. We verify the non-negativity of regular solutions (provided they exist) to the obtained macroscopic PDE-ODE system and perform numerical simulations to illustrate the solution behavior under several scenarios.
	\end{abstract}

\mysection{Introduction}\label{sec:intro}


The migration behavior of tumor cells under influence of biochemical and biophysical components of their environment is one of the cancer hallmarks \cite{Hanahan2011}. Glioma, one of the most common types of brain cancer, exhibits a high tendency to diffusive infiltration, thereby exploiting the anisotropy of brain tissue \cite{giese-etal96,Giese1996}. Gliomas in advanced stages (commonly called glioblastoma) develop large proportions of necrosis and are hypoxic, with exuberant angiogenic activity \cite{Brat2002,brat2003malignant}. The microscopic interplay of glioma cells among each other, with the surrounding structures, and with acidity (among other chemical cues) is decisive for the development and spread of the whole tumor. Understanding (some of) the complicated processes involved in the  evolution of a neoplasm can potentially help to improve therapy planning or even suggest new approaches. Here we propose a multiscale modeling approach to glioma invasion which connects single cell behavior with tumor scale dynamics.\\[-2ex]

Most of the available continuous models of glioma invasion are set exclusively on the macroscopic scale (for a recent review also addressing such settings see e.g. \cite{Alfonso17}), upon relying on simple flux balance, and many of them are versions or extensions of a model proposed by Murray some decades ago \cite{Murray1989}. Such reaction-diffusion systems have been further enlarged to include drift terms describing motility adjustment to extracellular signals: see e.g. \cite{Colombo2015,CTS20,Hogea2007,kim2009} for models explicitly dedicated to glioma, or the review in \cite{kolbe2020modeling} for settings with multiple taxis in the larger context of cell migration. Another modeling approach uses kinetic transport equations (KTEs) in the kinetic theory of active particles (KTAP) framework \cite{Bell-KTAP} to characterize the dynamics of distribution functions of densities of tumor cells sharing -supplementary to time and position- one or several kinetic variables (velocity and so-called activity variables). Among those models,  \cite{ConteSurulescu,CEKNSSW,Corbin2018,EHKS,EHS,EKS16,Hunt2016,kumar2020,PH13,Swan2017} refer to effects of brain tissue anisotropy on glioma invasion and deduce by macroscopic limits systems of reaction-(myopic)diffusion-taxis PDEs. Thereby, the taxis terms obtained in \cite{ConteSurulescu,CEKNSSW,Corbin2018,EHKS,EHS,EKS16,Hunt2016,kumar2020} are due to a multiscale approach which takes into account subcellular dynamics (receptor binding to soluble and insoluble components of the extracellular space), leading in the mesoscopic KTE to transport terms w.r.t. activity variables and turning rates depending on the same. The works \cite{Chauviere2007,HillenM5,kelkel2012multiscale,Lorenz2014,loy2019kinetic,Loy2020} address motility of eukaryotes in a heterogeneous environment, without specifically relating to glioma, but those models could also be employed to describe several migration aspects of this particular cell type. Still in this KTAP framework, alternatives leading on the macroscopic scale to various types of taxis are offered on the one hand in \cite{kumar2020,loy2019kinetic} by using 
turning rates depending on the pathwise gradient of some chemotactic signal, 
as originally proposed in \cite{othmer2002diffusion} for bacteria swimming, and on the other hand in \cite{Chauviere2007,CEKNSSW}, which consider cell stress and forces depending on the chemical and physical composition of the environment and acting on the cells, translating into transport terms w.r.t. the velocity variable in the corresponding KTE. In fact, the macroscopic limit of the KTE in \cite{CEKNSSW} led to a novel kind of haptotaxis, according to the dynamics of the mesoscopic tissue density depending on the local orientation of tissue fibers. \\[-2ex]

In the present note we propose an approach which is closely related to that in \cite{CEKNSSW}, however involves some differences in the description of single cell velocity dynamics (both speed and direction are varying) and in the way  we do the transition to the macroscopic level, on which a flux-saturated reaction-diffusion-taxis equation for the evolution of glioma cell density is obtained.\\[-2ex]

Flux limitations were considered increasingly often in connection with models describing cell motility, in order to alleviate the infinite speed of propagation triggered by linear diffusion and the excessive influence of the latter on the spread of cells. They can be encountered not only in the (nonlinear) diffusion part, but also in taxis terms, and reflect some kind of optimal transport in compliance of the respective population of cells to one or several tactic signals. While models directly including such terms on the macroscopic scale by a balance of fluxes were considered e.g. in \cite{CTS20,kim2009}, a careful derivation from KTEs has been provided formally in \cite{BBNS2010} and rigorously in \cite{perthame2018flux}. Both works were addressing cell chemotaxis, the former also obtaining flux-limiting self-diffusion. The deduction was achieved in both cases by an appropriate choice of the signal response function involved in the turning operator and depending on the directional derivative of the (chemotactic) signal. Here we propose an alternative approach which starts on the single cell scale by characterizing velocity dynamics, in particular having it influenced by spatial gradients of tissue, acidity, and isospecific cell densities. On the mesolevel this translates into a transport term w.r.t. the velocity variable, which  carries such gradients. By a formal macroscopic limit we deduce for the glioma cell density a PDE with flux-limited diffusion, chemo-, and haptotaxis.\\[-2ex] 

The rest of this paper is organized as follows: Section~\ref{sec:modeling} provides the set up of microscopic and mesoscopic dynamics of glioma cells and the macroscopic evolution of the factors in the tumor microenviroment which influence the development and spread of the neoplasm.  Section~\ref{sectionMacroscopic} contains the derivation of a fully macroscopic system featuring the interactions between glioma cell density, acidity, tissue, and vascularization. For the obtained model with flux-limited pH-taxis, self-diffusion, and haptotaxis, the non-negativity and upper bounds of regular enough solutions are proved, provided such solutions exist and the initial conditions satisfy analogous bounds. Numerical simulations are performed in Section~\ref{sec:numerics}. Eventually, Section~\ref{sec:discussion} provides a discussion of the this work's outcome, along with some perspectives.


\mysection{Multiscale modeling}\label{sec:modeling}

In this model the following aspects are to be taken into account:
\begin{itemize}
	\item migration of cancer cells due to pH gradients, tissue gradients and population pressure, incorporating the effects of tissue alignment,
	\item binding of cancer cells to tissue fibers,
	\item influence of acidic environment on tumor evolution,
	\item vascularization.
\end{itemize}

The multiscale modeling approach follows the ideas in several previous papers \cite{ConteSurulescu,CEKNSSW,Corbin2018,EHKS,EHS,EKS16,Hunt2016,kumar2020}. New in this note is the microscopic description of velocity dynamics - which is akin to that in \cite{CEKNSSW}, as it involves (signed) gradients of tactic signals, but here the cell speed is no longer constant and the cell density distribution influences the cell motility. The performed upscaling is related, however different from earlier limiting procedures and leads to a highly complex macroscopic PDE-ODE system featuring for glioma cell density self-diffusion and multiple taxis, all of which are flux-limited. 

\subsection{Microscopic scale}\label{subsec:micro}

\subsubsection{Dynamics of the receptor binding state $y$}\label{subsectionBindingDynamics} Let $R$ denote the amount of cell receptors which are able to bind to surrounding tissue. For simplicity we assume $R$ to be constant. The amount of free receptors on a cell in binding state $y$ is then given by $R-y$, with $y\in Y:=(0,R)$.  Let $k^+$ denote the attachment rate of a free receptor to adjacent tissue fibers, and let $k^-$ denote the corresponding detachment rate. Then the process of binding and unbinding in dependence on the macroscopic tissue density $Q(t,x)$ is described by
\begin{align*}
	(R-y)+\frac{Q}{K_Q}\xrightleftharpoons[k^{-}]{k^{+}}y,
\end{align*}
where the constant $K_Q>0$ represents the tissue carrying capacity. 
The corresponding ODE obtained by mass action kinetics is
\begin{align} \label{ydynamics}
	\dot{y}=k^+(R-y)\frac{Q}{K_Q}-k^-y=:G(Q,y).
\end{align}

\subsubsection{Dynamics of cell velocity $v$} \label{subsectionVelocityDynamics}
The migration of cancer cells is affected by different gradients. Increasing  gradients of acidity have a repelling effect, whereas the cells are attracted by gradients of tissue density. The smaller the amount of cell receptors bound to tissue, the more sensitive it reacts towards tissue gradients. We further assume that cancer cells try to avoid regions of high cell densities. Under these assumptions, the preferred direction of a cell can be modeled by a weighted sum of the gradients $-\nabla_x h$, $\nabla_x Q$ and $-\nabla_x M$, where $M$ represents the macroscopic tumor cell density. We choose 
$$b=(1-\rho_1-\rho_2)\frac{-\nabla h}{\sqrt{\left(\frac{K_h}{X}\right)^2+|\nabla h|^2}}+\rho_1\frac{R-y}{R}\frac{\nabla Q}{\sqrt{\left(\frac{K_Q}{X}\right)^2+|\nabla Q|^2}}+\rho_2\frac{-\nabla M}{\sqrt{\left(\frac{K_M}{X}\right)^2+|\nabla M|^2}},$$
where
$\rho_1, \rho_2\in (0,1)$ are constants and $X>0$ is also a constant to be selected in correspondence to appropriate time and length scales. We will address this issue in Subsection~\ref{subsec:nondimensionalization}.
Typically, glioma cells migrate along tissue fibers; they preferentially follow the white matter tracts consisting of bundles of such fibers \cite{Giese1996,GGritsenko2011}. Diffusion tensor imaging (DTI) provides a means to assess (with the aid of the water diffusion tensor $\mathbb{D}_W$) the anisotropic brain structure down to the level of voxels with edges of 1-2 mm.  The joint effect of fiber tract orientations and preferred direction relating to gradients leads to a change in velocity orientation of the form 
$$\mathbb{D}_Wb=\sum_{i=1}^N\alpha_i\omega_i\omega_i^Tb=\sum_{i=1}^N\alpha_i\omega_i\langle \omega_i,b \rangle,$$ 
where $\omega_i$ are normed eigenvectors of $\mathbb{D}_W$ with corresponding eigenvalues $\alpha_i$. The acceleration is then given by $$g(t,x)=a_1\frac{K_M-M}{K_M}\mathbb{D}_Wb,\, \ a_1>0,$$ 
where the factor $\frac{K_M-M}{K_M}$ is due to limited motility in crowded regions.\\[-2ex]

A cell which is not exposed to external signal gradients can slow down or move randomly, even in opposite direction. We model deceleration by a term $-a_2v$, $a_2>0$. Altogether we obtain the following equation for velocity dynamics:
\begin{align} \label{vdynamics}
	\frac{\partial v}{\partial t}=g(t,x)-a_2v=:S(v,y,h,Q,M).
\end{align}
We see that $g(t,x)$ is bounded:
\begin{align*}
	|g(t,x)|=|a_1\frac{K_M-M}{K_M}\mathbb{D}_Wb|= a_1\frac{K_M-M}{K_M}|\sum_{i=1}^N\alpha_i\omega_i\langle\omega_i,b\rangle|\leq a_1\alpha_{max}
\end{align*}
(the boundedness of $M$ by its carrying capacity $K_M$ will be shown in Subsection~\ref{subsec:bound-M}.).
Starting with speed $s:=|v|\leq s_{max}:=\frac{a_1}{a_2}\alpha_{max}$ and assuming the water diffusion tensor $\mathbb{D}_W$ to be constant in time, the speed $s_{max}$ cannot be exceeded. In case of a water diffusion tensor which varies in time and space, $\alpha_{max}$ and hence also $s_{max}$ depend on $t$ and $x$ and we have $s=|v|\leq \bar{s}_{max}:=\max\limits_{0\leq t\leq T,\ x\in\R^3}s_{max}(t,x)$.\\[-2ex]

For the cell positions we consider as usual the ODE system $\frac{dx}{dt}=v$. 

\subsection{Mesoscopic scale}\label{subsec:meso}

We consider the cell density function $p:[0,T]\times \R^N\times V\times Y\rightarrow \mathbb{R}^+$, $V\subset\mathbb{R}^N, Y\subset\mathbb{R}_0^+$, depending on time $t$, position $x$, velocity $v$, and activity variable $y$. The velocity vector $v=s\theta$ contains information on speed $s\in [0,s_{max}]$ and direction $\theta\in \mathbb S^{N-1}$ of a cell. The scalar variable $y$ denotes the amount of cell surface receptors bound to tissue. The macroscopic tumor cell density is obtained by averaging over all velocities and all activity variables:
$$M(t,x)=\int _Y\int _V p(t,x,v,y)dv\ dy.$$
Then the dynamics of $p$ can be  described by way of a kinetic transport equation of the form 
\begin{align} \label{mesoscopic}
	\frac{\partial p}{\partial t}+\nabla_x\cdot(vp)+\partial_y(G(Q,y)p)+\nabla_v\cdot(S(v,y,h,Q,M)p)=\beta(p),
\end{align}
where the right hand side $\beta (p)$ accounts for source terms (proliferation) to be addressed below. This is another difference to previous models \cite{Chauviere2007,ConteSurulescu,CEKNSSW,Corbin2018,EHKS,EHS,EKS16,HillenM5,Hunt2016,PH13} in the kinetic theory of active particles (KTAP) framework \cite{Bell-KTAP}, where the right hand side usually describes velocity reorientations by way of a turning operator in integral form.\\[-2ex]

The proliferative activity of cancer cells depends on their actual binding state. Without connection to the surrounding tissue, cells cannot perform mitosis and even die through anoikis \cite{anoikis.1994,Liotta2004}. On the other hand, too many bounds also inhibit cell division. We will factorize the proliferation rate into a part $\mu_1$, which is independent of $y$, and a part $\mu_2,$ which depends on $y$ and for which we choose $\mu_2(y)=\frac{y(R-y)}{R^2}$. Therewith, the proliferation is nearly turned off when there are too less or too many receptors bound to tissue. The $y$-independent part of the proliferation rate is modeled due to the assumption of glioma cells not being able to proliferate and migrate at the same time, also known as go-or-grow dichotomy \cite{giese-etal96,Zheng2009}. Unlike previous models \cite{ConteSurulescu,EKS16,Hunt2016,kolbe2020modeling,stinner-surulescu-uatay,ZSH} where the tumor cells are split into mutually exclusive migrating and proliferative subpopulations, the mentioned dichotomous behavior is taken here into account only by relating the $y$-independent part of the proliferation rate to cell speed in a decreasing manner. As the adaptation of speed to the surrounding environment happens fast compared with the time needed for proliferation, we approximate the velocity by the quasi-steady state $v^*$ of its dynamics. The corresponding speed is denoted by $s^*=|v^*|$. Upon also taking into account the detrimental  influences of a highly acidic environment as well as of population pressure by surrounding cancer cells, we propose for the $y$-independent part of the proliferation rate
$$\mu_1(M,h,s^*)=\mu\frac{s_{max}-s^*}{s_{max}}\left(1-\frac{M}{K_M}\right)\frac{K_h}{K_h+h},$$
with $\mu , K_h>0$ being two constants, the latter representing a threshold acidity level beyond which the cancer cells cannot advance trough the cell cycle leading to mitosis \cite{Vaupel,Webb2011}.  
After proliferation, the binding state of the daughter cells might differ from the original state. \\[-2ex]

Assuming that the receptor binding states of daughter cells are distributed symmetrically around the quasi-steady state $y^*$ of~\eqref{ydynamics}, i.e. $\int _Y(y-y^*)\chi (t,x,y,y')dy'=0$, and that they do not depend on the original activity states of the mother cells, we are led to choosing
\begin{align*}
	\beta(p)=\mu_1(M,h,s^*)\int_Y\mu_2(y')\chi(t,x,y)p(t,x,v,y')\dd y',
\end{align*}
where $\chi$ is a probability kernel representing the likelihood of cells to receive a receptor binding regime $y$ after division. As such, it holds that $\int _Y\chi(t,x,y)dy=1$. We also assume here that the activity-dependent component $\mu_2$ of the proliferation rate does only depend on the receptor binding regime available at the initiation of mitosis.

\subsection{Macroscopic scale}\label{subsec:macro}

\subsubsection{Tissue}
The acidity produced by the tumor cells by upregulated glycolysis degrades the surrounding tissue. Assuming that the latter is regenerated in a logistic way, we take
\begin{align}\label{tissue}
	\partial_tQ=c_1 Q\left(1-\frac{Q}{K_Q}-\frac{M}{K_M}\right)-c_2 \frac{h}{K_h+h}Q,
\end{align}
with $c_1, c_2>0$ constants. The constant $K_h>0$ has the same significance as above in $\mu_1$. 
For the initial condition we choose 
\begin{equation}\label{tissue_ini}
Q(0,x)=K_Q\left(1-\sqrt{\frac{tr(\mathbb{D}_W(x))}{3d_{ref}}}\right),
\end{equation}
 where the constant $d_{ref}$ is the maximum value (taken over all positions $x$) any of the entries of $\mathbb{D}_W$ can reach (corresponding to the diagonal entries of $\mathbb{D}_W$ for no surrounding tissue). \footnote{Recall that $\mathbb{D}_W(x)$ assesses the diffusivity of water molecules in a voxel with center at $x$, which is highest when the tissue -if available- is perfectly aligned, i.e. when there are two zero eigenvalues and the third, dominant eigenvalue dictates the local orientation.}

\subsubsection{Acidity and vascularization}\label{subsec:acid-vasc}
The dynamics of acidity concentration $h$ in the tumor microenvironment is modeled by
\begin{align} \label{protons}
	\partial_t h=D_h\Delta h+\gamma\frac{M}{K_M+M}\left(1-\frac{h}{K_h}\right)_+-\delta he,
\end{align}
where the second term on the right hand side describes proton production by tumor cells which is limited by the acidity threshold $K_h$, whereas the third term describes uptake by blood vessels which are represented by the density $e$ of endothelial cells. In fact, it can be shown that all solutions $h$ of~\eqref{protons} stay nonnegative and never exceed $K_h$ (if $0\le h(0,x)\le K_h$), so that the second term on the right hand side can be taken without the positive part of the parenthesis therein.\\[-2ex]

The tumor itself stimulates growth of blood vessels by producing certain growth factors. The latter are increasingly expressed when the cancer cell environment becomes hypoxic; this is typically occurring at sites with high tumor cell density. Since we do not want to inflate the model with yet another space-time dependent variable explicitly accounting for the concentration of such growth factor, we propose instead a chemotactic bias of endothelial cells towards regions with lower pH and choose for their evolution
\begin{align}\label{vascularization}
	\partial_t e=D_e\Delta e-\varsigma_e\nabla\cdot \left(e\left(1-\frac{e}{K_e}\right)\nabla h\right)+G_e(h,M)e\left(1-\frac{e}{K_e}\right).
\end{align}
The growth term $G_e(h,M)$ should be increasing w.r.t. $h$ and $M$, and could be assigned e.g., the form $G_e(h,M)=\mu _e\frac{hM}{K_hK_M+hM}$. Moreover, we assume that the tactic sensitivity is decreasing with the amount of available vasculature.

\subsection{Non-dimensionalization} \label{subsec:nondimensionalization}
Before deducing a macroscopic model, we non-dimensionalize equations (\ref{mesoscopic})-(\ref{vascularization}). To this aim, we define 
\begin{align*}
	\hat t=\frac{t}{\tau},\ \hat x=\frac{x}{X}, \ \hat{y}=\frac{y}{R},\;\hat{v}=\frac{v}{s_{max}},\;\hat{p}=\frac{Rs_{max}}{K_M}p,\;\hat{Q}=\frac{Q}{K_Q},\;\hat{h}=\frac{h}{K_h},\; \hat{e}=\frac{e}{K_e},\; \hat{M}=\iint \hat{p}\dd \hat{v}\dd\hat{y}.
\end{align*}
Note that (with $\hat s=s/s_{max}$) $$\hat{M}=\int _0^1\int _{\mathbb S^{N-1}}\int _0^1 \hat{p}\ \dd \hat{y}\dd \theta \dd \hat s=\iint _{V\times Y}\frac{Rs_{max}}{K_M}p\cdot \frac{1}{Rs_{max}}\dd (v,y)=\frac{M}{K_M}.$$
Doing the above transformations on the terms of (\ref{mesoscopic}) and multiplying the outcome by $\frac{R\tau s_{max}}{K_M}$ we arrive at
\begin{align}\label{mesoscopic_nondim1}
	\partial_{\hat t}\hat{p}+\nabla_{\hat x}\cdot\left(\hat{v}\hat{p}\right) +\tau k^-\partial_{\hat y}\left (\hat{G}(\hat Q,\hat{y})\hat{p}\right )+a_2\tau \nabla_{\hat{v}}\cdot \left(\hat{S}(\hat{v},\hat{y},\hat h,\hat Q, \hat M)\hat{p}\right)=\mu \tau \hat \beta (\hat p),
\end{align}
where we took $X=s_{max}\tau$ and 
where
\begin{align*}
	\hat{G}(\hat Q,\hat{y})&=\hat \kappa (1-\hat{y})\hat{Q}-\hat{y},\quad \text{with}\quad \hat \kappa:=\frac{k^+}{k^-}\\
	\hat{S}(\hat{v},\hat{y},\hat h, \hat Q,\hat M)&=\frac{a_1}{a_2s_{max}}(1-\hat{M})\mathbb{D}_W\hat b-\hat v,
\end{align*}
with
\begin{align*}
	&\hat b=(1-\rho_1-\rho_2)\frac{-\nabla \hat{h}}{\sqrt{1+|\nabla \hat{h}|^2}}+\rho_1(1-\hat{y})\frac{\nabla \hat{Q}}{\sqrt{1+|\nabla \hat{Q}|^2}}+\rho_2\frac{-\nabla \hat{M}}{\sqrt{1+|\nabla \hat{M}|^2}},\\
	&\hat{\beta}(\hat{p})=(1-\hat{M})\hat{\eta}\int_0^1\hat{y}'(1-\hat{y}')\hat{\chi}(\hat{y})\hat{p}(\hat{y}')\dd \hat{y}',\;\hat{\eta}(\hat{h},\hat{s}^*)=\frac{1-\hat s^*}{1+\hat{h}},\;\hat s^*=\frac{s^*}{s_{max}},\; \hat{\chi}(\hat{y})=R\chi(R\hat y).
\end{align*}
Note that $\int_0^1\hat{\chi}(\hat{y})\dd \hat{y}=\int_0^1R\chi(R\hat y)\dd \hat{y}=\int_0^1R\chi(y)\frac1R\dd y=1$.
\\[-2ex]

Equation (\ref{tissue}) is rescaled as 
\begin{align} \label{tissue_half-nondimensionalized}
	\partial_{\hat t}\hat{Q}=\hat c_1\hat{Q}(1-\hat{Q}-\hat M)-\hat c_2\frac{\hat{h}}{1+\hat{h}}\hat{Q},
\end{align}
with $\hat c_i=c_i\tau $ ($i=1,2$) and the initial condition becoming 
\begin{align*}
	\hat{Q}(0,\hat x)=1-\sqrt{\frac{tr(\mathbb{D}_W(\hat x))}{3d_{ref}}}.
\end{align*}
From (\ref{protons}) we obtain 
\begin{align} \label{protons_half-nondimensionalized}
	\partial_{\hat t}\hat{h}=\hat D_h\Delta\hat{h}+\hat{\gamma}(1-\hat{h})\frac{\hat{M}}{1+\hat{M}}-\hat{\delta}\hat{h}\hat{e},
\end{align}
where $\hat D_h=\frac{D_h\tau }{X^2}=\frac{D_h}{\tau s_{max}^2}$, $\hat{\gamma}=\frac{\gamma \tau }{K_h},\,\hat{\delta}=\delta \tau K_e$.
Finally, we obtain from (\ref{vascularization})
\begin{align} \label{vascularization_half-nondimensionalized}
	\partial_{\hat t}\hat{e}=\hat D_e\Delta\hat{e}-\hat{\varsigma}_e\nabla\cdot\left(\hat{e}(1-\hat{e})\nabla\hat{h}\right)+\hat{G}_e(\hat{h},\hat{M})\hat{e}(1-\hat{e}),
\end{align}
where $\hat D_e=\frac{D_e}{\tau s_{max}^2}$,  $\hat{\varsigma}_e=\frac{\varsigma_eK_h}{\tau s_{max}^2},\,\hat{G}_e(\hat{h},\hat{M})=\mu_e\frac{\tau \hat h\hat M}{1+\hat h\hat M}$.\\[-2ex]

\noindent
In the following we will drop the hat symbol from all variables, for simplicity of writing. We are still free to choose the scaling constant $\tau $ and set $\tau :=1/\mu $, which means that our typical time corresponds to the (average) proliferation time of glioma cells. Thus, we obtain the nondimensonalized system 
\begin{subequations}\label{nondimensionalized_system}
	\begin{align}
		&\partial_tp+\nabla_x\cdot\left(vp\right) +\frac{k^-}{\mu}\partial_y\left (G(Q,y)p\right )+\frac{a_2}{\mu}\nabla_v\cdot \left(S(v,y,h,Q,M)p\right)=\beta (p),\label{nondimensionalized_system-p}\\
		&\partial_tQ=c_1Q(1-Q-M)-c_2\frac{h}{1+h}Q,\label{nondimensionalized_system-Q}\\
		&\partial_th=D_h\Delta h+\gamma (1-h)\frac{M}{1+M}-\delta he,\label{nondimensionalized_system-h}\\
		&\partial_te=D_e\Delta e-\varsigma_e\nabla\cdot\left(e(1-e)\nabla h\right)+G_e(h,M)e(1-e),\label{nondimensionalized_system-e}\\
		&\mbox{with}\nonumber\\
		&G(Q,y)=\kappa (1-y)Q-y,\label{nondimensionalized_system-G}\\
		&S(v,y,h,Q,M)=\frac{a_1}{a_2s_{max}}(1-M)\mathbb{D}_Wb-v,\label{nondimensionalized_system-S}	\\
		&b=	(1-\rho_1-\rho_2)\frac{-\nabla h}{\sqrt{1+|\nabla h|^2}}+\rho_1(1-y)\frac{\nabla Q}{\sqrt{1+|\nabla Q|^2}}+\rho_2\frac{-\nabla M}{\sqrt{1+|\nabla M|^2}},\label{nondimensionalized_system-b}\\
		&\beta(p)=(1-M)\eta(h,s^*)\int_0^1y'(1-y')\chi(y)p(y')\dd y',\quad\eta(h,s^*)=\frac{1-s^*}{1+h},\label{nondimensionalized_system-beta-eta}\\
		&G_e(h,M)=\nu_e\frac{hM}{1+hM},\quad \nu_e:=\frac{\mu_e}{\mu}.\label{nondimensionalized_system-G_e}
	\end{align}
\end{subequations}

The kinetic equation~\eqref{nondimensionalized_system-p} is still characterizing mesoscopic dynamics of cancer cells, as $p$ depends on time, position, velocity, and the activity variable (amount of receptors bound to tissue fibers). Thus, the attempt to solve system ~\eqref{nondimensionalized_system} numerically has to face the high dimensionality of the phase space $\R^N\times \left ((0,1)\times \mathbb S^{N-1}\right )\times (0,1)$, which is quite inconvenient. Therefore, in the next section we aim at deducing a macroscopic counterpart of~\eqref{nondimensionalized_system-p}, to be coupled with the rest of equations in~\eqref{nondimensionalized_system}.

\section{Derivation of a fully macroscopic system} \label{sectionMacroscopic}
\subsection{Assumptions and notations}\label{subsec:assumptions} We make the following simplifying assumptions, which will be needed in the process of obtaining a closed system by integrating w.r.t. $y$ and $v$:
\begin{align*}
	\int_V\int_Y(v-v^*)(y-y^*)p\dd y\dd v&\approx0,&&\int_V\int_Y(y-y^*)^2p\dd y\dd v\approx0,\\ 
	\int_V\int_Y(v_i-v_i^*)(y-y^*)^2p\dd y\dd v&\approx0 
	\quad\mbox{and}\quad &&\nabla_x\cdot\int_V\int_Y(v_i-v_i^*)(v-v^*)p\dd y\dd v\approx0,
\end{align*}
where $v_i$ is the i-th component of the vector $v$  and $y^*=\frac{Q}{Q+1/\kappa }$ and $v^*=\frac{a_1}{a_2s_{max}}(1-M)\mathbb{D}_Wb$ are the quasi-stationary states of the correspondingly nondimensionalized microscopic dynamics (\ref{ydynamics}) and (\ref{vdynamics}). Thus, we assume that some of the second order moments for the tumor cell distribution w.r.t. deviations of $v$ and $y$ from their steady-states are negligible, which is reasonable, since the microscopic dynamics of receptor binding and velocity innovations happen very fast in comparison to the (mesoscopic) behavior  of cell groups sharing the same regimes of activity and kinetic variables. Likewise, the third order moment involving $(y-y^*)^2$ vanishes. The (partial) second order moment w.r.t. $v$ is not required to nullify, but only its divergence.\\[-2ex]

Subsequently we use the following notations:
\begin{align}
	&M(t,x):=\int_V\int_Yp\dd y\dd v,\quad M^y(t,x):=\int_V\int_Y yp\dd y\dd v,\quad M^v_i(t,x):=\int_V\int_Y v_ip\dd y\dd v,\notag \\ &M^v(t,x):=\int_V\int_Y vp\dd y\dd v=\left(M^v_i\right)_{i=1}^N.\label{eq:def-moments}
\end{align}

\subsection{Boundary conditions w.r.t. kinetic variables}
Due to the performed non-dimensionalization, the domains $Y$ and $V$ are given by $$Y=(0,1) \;\mbox{and}\; V=B_{1}^N(0)=(0,1)\times \mathbb S^{N-1}.$$ As in earlier works \cite{Chauviere2007,EHKS,EHS,EKS16,kelkel2012multiscale} we assume $p$ to be compactly supported in the $V\times Y$ space. 

\begin{remark}
	Equation (\ref{nondimensionalized_system-p}) is of transport type with respect to $y$ and $v$. Hence, boundary conditions w.r.t. these variables need only be prescribed at the inflow boundary of $Y$ and $V$.
	\begin{itemize}
		\item Inflow boundary of $Y$: The dynamics of $y$ is given by $\dot y=G(Q,y)$, with the right hand side~\eqref{nondimensionalized_system-G}. A binding state $y\in \partial Y$ is part of the inflow boundary if $G(Q,y)\cdot n\le 0$, where $n$ is the outward normal on the boundary. On $\partial Y=\{0,1\}$ it holds $$G(Q,0)\cdot n(0)=\kappa Q\cdot(-1)\le 0\;\mbox{and}\;G(Q,1)\cdot n(1)=-1<0.$$ Hence, the inflow boundary of $Y$ coincides with $\partial Y$. Thus, boundary conditions can be prescribed on the whole of $\partial Y$.
		\item Inflow boundary of $V$: The dynamics of $v$ is determined by $\dot v=S(v,y,h,Q,M)$ with the right hand side~\eqref{nondimensionalized_system-S}.  Now let $v\in\partial V,$ so $|v|=1$. The corresponding outward normal vector is then given by $n=v$, and we obtain
		\begin{align*}
			S(v,y,h,Q,M)\cdot n&=\left\langle \frac{a_1}{a_2s_{max} }(1-M)\mathbb{D}_Wb,v\right\rangle-\left\langle v,v\right\rangle\\&=\frac{a_1}{a_2s_{max} }(1-M)\left\langle \sum_{i=1}^N\alpha_i\omega_i\left\langle \omega_i,b\right\rangle,v\right\rangle-|v|^2\\ &\leq\frac{a_1}{a_2s_{max} }(1-M)\left|\sum_{i=1}^N\alpha_i\omega_i\left\langle \omega_i,b\right\rangle\right|-1\\&\leq \frac{a_1}{a_2s_{max} }\alpha_{max}\underbrace{|b|}_{<1}-1\\&<\frac{a_1}{a_2 s_{max}}\frac{a_2}{a_1} s_{max}-1=0.
		\end{align*}
		Hence, $V$ only has an inflow boundary, therefore boundary conditions can be prescribed on the whole of $\partial V$.
	\end{itemize}
\end{remark}


\subsection{Equations for the moments~\eqref{eq:def-moments}}
Let us integrate (\ref{nondimensionalized_system-p}) with respect to $y$ and $v$:
\begin{align*}
	\partial_tM&+\nabla_x\cdot M^v+\frac{k^-}{\mu}\int_V\int_Y\partial _y(G(Q,y)p)\dd y\dd v+\frac{a_2}{\mu }\int_V\int_Y\nabla_v\cdot(S(v,y,h,Q,M)p)\dd y\dd v\\
	&=\int_V\int_Y\beta(p)\dd y\dd v.
\end{align*}
The third and fourth term on the left hand side are zero due to the chosen boundary conditions. For the integral on the right hand side we find
\begin{align*}
	\int_V\int_Y\beta(p)\dd y\dd v&=\int_V\int_Y(1-M)\eta(h,s^*)\int_Yy'(1-y')\chi(y)p(y')\dd y'\dd y\dd v\\
	&=(1-M)\eta(h,s^*)\underbrace{\int_V\int_Yy'(1-y')p(y')\dd y'\dd v}_{(A)}
\end{align*}
\begin{align*}
	(A)&=\int_V\int_Yy(1-y)p(y)\dd y\dd v=\int_V\int_Yyp(y)\dd y\dd v-\int_V\int_Yy^2p(y)\dd y\dd v\\
	&=M^y-\underbrace{\int_V\int_Y(y-y^*)^2p(y)\dd y\dd v}_{\approx 0}-\int_V\int_Y2y^*yp(y)\dd y\dd v+\int_V\int_Y(y^*)^2p(y)\dd y\dd v\\
	&=M^y-2y^*M^y+(y^*)^2M
\end{align*}
\begin{align*}
	\Rightarrow\quad\int_V\int_Y\beta(p)\dd y\dd v=(1-M)\eta(h,s^*)\left(M^y-2y^*M^y+(y^*)^2M\right).
\end{align*}
Hence, we obtain the macroscopic equation
\begin{align}\label{macroM}
	\partial_tM+\nabla_x\cdot M^v=\eta(h,s^*)(1-M)\left(M^y-2y^*M^y+(y^*)^2M\right).
\end{align}
To obtain a closed system we need further equations, for the moments $M^y$ and $M^v$. To this aim, we multiply (\ref{nondimensionalized_system-p}) by $y$ and integrate again with respect to $y$ und $v$:
\begin{align}
	\partial_tM^y&+\nabla_x\cdot\int_Y\int_Vvyp\dd y\dd v+\frac{k^-}{\mu}\int_V\int_Yy\partial _y(G(Q,y)p)\dd y\dd v\notag \\
	&+\frac{a_2}{\mu }\int_V\int_Yy\nabla_v\cdot(S(v,y,h,Q,M)p)\dd y\dd v
	=\int_V\int_Yy\beta(p)\dd y\dd v.\label{way-to-M^y}
\end{align}
Again, the fourth term is zero due to the chosen boundary conditions. The third term on the left hand side can be computed by partial integration:
\begin{align*}
	\frac{k^-}{\mu}\int_V\int_Yy\partial _y(G(Q,y)p)\dd y\dd v&=-\frac{k^-}{\mu}\int_V\int_YG(Q,y)p\dd y\dd v=-\frac{k^-}{\mu}\int_V\int_Y(\kappa Q(1-y)-y)p\dd y\dd v\\
	&=\frac{k^-}{\mu }(\kappa Q+1)M^y-\frac{k^-\kappa }{\mu }QM.
\end{align*}
For the remaining terms we find
\begin{align*}
	\nabla_x\cdot\int_V\int_Yvyp\dd y\dd v&=\nabla_x\cdot\int_V\int_Y(v-v^*)(y-y^*)p\dd y\dd v+\nabla_x\cdot\int_V\int_Y(vy^*+v^*y)p\dd y\dd v\\&\quad-\nabla_x\cdot\int_V\int_Yy^*v^*p\dd y\dd v\\
	&=\nabla_x\cdot(y^*M^v+v^*M^y-y^*v^*M),
\end{align*}
\begin{align*}
	\int_V\int_Yy\beta(p)\dd y\dd v&=\eta(h,s^*)(1-M)\int_V\int_Yy\chi(y)\dd y\int_Yy'(1-y')p(y')\dd y'\dd v\\
	&=\eta(h,s^*)(1-M)y^*(M^y-2y^*M^y+(y^*)^2M),
\end{align*}
where we used the symmetry of $\chi$ around $y^*$:
\begin{align*}
	\int_Yy\chi(y)\dd y=\underbrace{\int_Y(y-y^*)\chi(y)\dd y}_{=0}+y^*\underbrace{\int_Y\chi(y)\dd y}_{=1}=y^*.
\end{align*}
Putting the above terms together, we find from~\eqref{way-to-M^y}
\begin{align}\label{macroMy}
	\begin{split}
		\partial_tM^y+&\nabla_x\cdot(y^*M^v+v^*M^y-y^*v^*M)+\frac{k^-}{\mu }(\kappa Q+1)M^y-\frac{k^-\kappa }{\mu }QM\\
		&=\eta(h,s^*)(1-M)y^*(M^y-2y^*M^y+(y^*)^2M).
	\end{split}
\end{align}
To find an equation for $M^v$, we repeat the computations from above, now multiplying (\ref{nondimensionalized_system-p}) by $v_i$ instead of $y$. 
Integration w.r.t. $v$ and $y$ yields
\begin{align*}
	\partial_tM_i^v+&\nabla_x\cdot\int_V\int_Yv_i vp\dd y\dd v+\frac{ a_2}{\mu}\int_V\int_Yv_i\nabla_v\cdot(S(v,y,h,Q,M)p)\dd y\dd v=\int_V\int_Yv_i\beta(p)\dd y\dd v.
\end{align*}
We compute the terms separately:
\begin{align*}
	\nabla_x\cdot\int_V\int_Yv_ivp\dd y\dd v&=
	\nabla_x\cdot\int_V\int_Y(v_i-v_i^*)(v-v^*)p\dd y\dd v + \nabla_x\cdot\int_V\int_Y(v_iv^*+v_i^*v-v_i^*v^*)p\dd y\dd v\\
	&=\nabla_x\cdot(v^*M_i^v+v_i^*M^v-v_i^*v^*M).
\end{align*}
For simplicity of writing we will use the notation $S(v,y):=S(v,y,h,Q,M)$, but keep in mind the dependency on the macroscopic quantities $h,Q,M$. We compute
\begin{align*}
	\int_V\int_Yv_i\nabla_v\cdot(S(v,y)p)\dd y\dd v&= \int_Y\left[\int_Vv_i\partial_{v_i}(S_i(v,y)p)\dd v+\sum_{\substack{j=1,j\neq i}}^N\int_Vv_i\partial_{v_j}(S_j(v,y)p)\dd v\right]\dd y\\
	&=\int_Y\int_{V_{\neq i}}\int_{V_i}v_i\partial_{v_i}(S_i(v,y)p)\dd v_i\dd\tilde{v}\dd y\\ &\quad+ \sum_{\substack{j=1,j\neq i}}^N\int_Y\int_{V_{\neq j}}v_i\int_{V_j}\partial_{v_j}(S_j(v,y)p)\dd v_j\dd\tilde{v}\dd y\\
	&= \int_Y\int_{V_{\neq i}}\left( v_i\underbrace{\left.S_i(v,y)p\right|_{\partial V_i}}_{=0} -\int_{V_i}S_i(v,y)p\dd v_i\right)\dd \tilde{v}\dd y\\ 
	&=-\int_Y\int_VS_i(v,y)p\dd v\dd y=-\int_Y\int_V(\tilde{g}_ip+y\tilde{\tilde g}_ip-v_ip)\dd v\dd y\\
	&=-\tilde{g}_iM-\tilde {\tilde g}_iM^y+M_i^v,
\end{align*}
where we used the notation $v=(v_i,\tilde v)\in V_i\times V_{\neq i}=V$, along with (recall~\eqref{nondimensionalized_system-S}))
\begin{align*}
	&S(v,y)=g(y)-v=\tilde{g}+y\tilde{\tilde g}-v,\\
	&\tilde{g}:=\frac{a_1}{a_2s_{max} }(1-M)\mathbb{D}_W \left((1-\rho_1-\rho_2)\frac{-\nabla h}{\sqrt{1+|\nabla h|^2}}+\rho_1\frac{\nabla Q}{\sqrt{1+|\nabla Q|^2}}+\rho_2\frac{-\nabla M}{\sqrt{1+|\nabla M|^2}}\right),\\
	&\tilde{\tilde g}:=-\frac{a_1}{a_2s_{max} }\rho_1(1-M)\mathbb{D}_W\frac{\nabla Q}{\sqrt{1+|\nabla Q|^2}}.
\end{align*}
Eventually,
\begin{align*}
	\int_V\int_Yv_i\beta(p)\dd y\dd v&=\eta(h,s^*)(1-M)\int_V\int_Yv_i\int_Y\chi(y)y'(1-y')p(y')\dd y'\dd y\dd v\\
	&=\eta(h,s^*)(1-M)\int_V\int_Yv_iy'(1-y')p(y')\dd y'\dd v\\
	&=\eta(h,s^*)(1-M)\left (\int_V\int_Y(v_i-v_i^*)(y-y^2)p(y)\dd y\dd v+ v_i^*\int_V\int_Y(y-y^2)p(y)\dd y\dd v\right )\\
	&=\eta(h,s^*)(1-M)\Bigg (\int_V\int_Y(v_i-v_i^*)(y-y^*)p(y)\dd y\dd v+y^*\int_V\int_Y(v_i-v_i^*)p(y)\dd y\dd v\\&\quad -\int_V\int_Y(v_i-v_i^*)(y-y^*)^2p(y)\dd y\dd v-\int_V\int_Y(v_i-v_i^*)(2yy^*-(y^*)^2)p(y)\dd y\dd v\\&\quad +\int_V\int_Yv_i^*yp(y)\dd y\dd v-\int_V\int_Yv_i^*(y-y^*)^2p(y)\dd y\dd v\\
	&\quad -\int_V\int_Yv_i^*(2yy^*-(y^*)^2)p(y)\dd y\dd v\Bigg )\\
	&=\eta(h,s^*)(1-M)\Bigg ( y^*M_i^v- y^*v_i^*M-\int_V\int_Yv_i(2yy^*-(y^*)^2)p(y)\dd y\dd v+ v_i^*M^y\Bigg )\\
	&=\eta(h,s^*)(1-M)\Bigg (v_i^*(2y^*-1)(y^*M-M^y)+y^*(1-y^*)M_i^v\Bigg ),
\end{align*}
where we used $\int_V\int_Yv_iyp\dd y \dd v=\int_V\int_Y\left (v_iy-(v_i-v_i^*)(y-y^*)\right)p\dd y \dd v$, in virtue of our assumptions in Subsection~\ref{subsec:assumptions}.

Hence, summarizing the terms calculated above, we find 
\begin{align}\label{macroMv}
	\begin{split}
		\partial_tM_i^v&+\nabla_x\cdot \left(v^*M_i^v+v_i^*M^v-v_i^*v^*M\right)+\frac{a_2}{\mu } \left(M_i^v-\tilde{g}_iM-\tilde{\tilde g}_iM^y\right)\\
		&=\eta(h,s^*)(1-M)\Big (v_i^*(2y^*-1)(y^*M-M^y)+y^*(1-y^*)M_i^v\Big ).
	\end{split}
\end{align}
for $i=1,2,...,N$. Together, (\ref{macroM}),(\ref{macroMy}) and (\ref{macroMv}) form a closed macroscopic system.

\subsection{Upscaling}\label{subsec:upscaling}

The aim of this subsection is to derive a single macrosopic equation for $M$ from the system (\ref{macroM}) - (\ref{macroMv}) by scaling methods. For this we take a closer look at the involved parameters.  
In literature, the following values  can be found:
\begin{itemize}
	\item $s_{max}\sim 0.8-1\frac{\mu m}{min}$\quad  \cite{milo2015cell,Prag283};
	\item $\alpha_{max}\sim 12\cdot 10^4\frac{\mu m^2}{min}$\quad  \cite{Thomsen1987} ($9\cdot 10^4\frac{\mu m^2}{min}$ in white matter, $3\%$ SD; $13.8\cdot 10^4\frac{\mu m^2}{min}$ in grey matter, $7\%$ SD);
	\item $\mu\sim10^{-5}-1.5\cdot 10^{-5}\frac{1}{sec}=6\cdot 10^{-4}-9\cdot 10^{-4}\frac{1}{min}$ \cite{Weber2014}; this is in agreement with the values provided for $\tau$ in \cite{Chaplain2006}; 
	\item $k^-\sim 0.6\frac{1}{min}$\quad  \cite{EHKS,Lauffenburger}
\end{itemize}
There does not seem to be reliable data on $a_1$ (with units $\frac{1}{\mu m\cdot min}$), which is the parameter scaling cell acceleration, thus we can so far estimate $$a_2=\frac{a_1\alpha_{max}}{s_{max}}\sim 12a_1\cdot 10^4\frac{1}{min}.$$ 
Setting 
$$\epsilon:=\frac{\mu}{a_2}\approx \frac{5}{a_1}10^{-9},$$ 
this is a very small number, no matter what (reasonable) value $a_1$ takes. 
We estimate $\epsilon \sim O(10^{-3})$ (at most, rather smaller, in virtue of the tiny masses and stresses of cells). 
On the other hand we also have $$\frac{\mu}{k^-}\approx \epsilon,$$ which motivates to set $\tau =1/\epsilon$, hence the time is scaled by $\epsilon$. Our choice of the typical length $X=s_{max} \tau$ suggests that we should have the same $\epsilon$-scaling for the space variable.\\[-2ex] 

Applying these estimates to our equations~\eqref{macroM},~\eqref{macroMy}, ~\eqref{macroMv}) deduced above, we get
\begin{align} \label{scaled_M}
	&\partial_tM+\nabla_x\cdot M^v=\eta(h,s^*)(1-M)\left(M^y-2y^*M^y+(y^*)^2M\right),\\
	&\epsilon\partial_tM^y+\epsilon\nabla_x\cdot(y^*M^v+v^*M^y-y^*v^*M)+(\kappa Q+1)M^y-\kappa QM\notag \\
	&\qquad \qquad\qquad \qquad =\epsilon\eta(h,s^*)(1-M)y^*(M^y-2y^*M^y+(y^*)^2M),\label{scaled_My}\\
	&\epsilon\partial_tM_i^v+\epsilon\nabla_x\cdot \left(v^*M_i^v+v_i^*M^v-v_i^*v^*M\right)+M_i^v-\tilde{g}_iM-\tilde {\tilde g}_iM^y\notag \\
	&\qquad \qquad\qquad \qquad=\epsilon\eta(h,s^*)(1-M)\Big (v_i^*(2y^*-1)(y^*M-M^y)+y^*(1-y^*)M_i^v\Big ).\label{scaled_Mv}
\end{align}

We consider Hilbert expansions for the moments:
\begin{align*}
	&M=M_0+\epsilon M_1+...\\
	&M^v=M_0^v+\epsilon M_1^v+...,\\
	&M^y=M_0^y+\epsilon M_1^y+...
\end{align*}
in (\ref{scaled_M})-(\ref{scaled_Mv}) and sort by orders of $\epsilon,$ considering only the leading order terms.\\

From~\eqref{scaled_My} we have
\begin{align}\label{M0y}
	(\kappa Q+1)M_0^y=\kappa QM_0\;\;\Rightarrow\;\;M_0^y=\frac{\kappa Q}{\kappa Q+1}M_0=y^*M_0.
\end{align} Equation (\ref{scaled_Mv}) yields
\begin{align*}
	M_{0,i}^v-\tilde{g}_iM_0-\tilde{\tilde g}_iM_0^y=0,
\end{align*}
where $\tilde g_i=\tilde g_i(M_0)$. Using (\ref{M0y}) we find
\begin{align}\label{M0v}
	M_{0,i}^v=(\tilde{g}_i+y^*\tilde {\tilde g}_i)M_0=g_i(y^*)M_0.
\end{align}

Collecting leading order terms in (\ref{scaled_M}) and using (\ref{M0y}) and (\ref{M0v}), we find
\begin{align}
	\partial_tM_0+\nabla_x\cdot\left(g(y^*)M_0\right)=\eta(h,s^*)\left(y^*-(y^*)^2\right)M_0(1-M_0),
\end{align}
where 
\begin{subequations}\label{coeff-M-eq-final}
	\begin{align}
		&g(y^*)=\frac{a_1}{a_2 s_{max}}(1-M_0)\mathbb{D}_Wb(y^*),\label{eq:g}\\ 
		&b(y^*)=(1-\rho_1-\rho_2)\frac{-\nabla h}{\sqrt{1+|\nabla h|^2}}+\rho_1(1-y^*)\frac{\nabla Q}{\sqrt{1+|\nabla Q|^2}}+\rho_2\frac{-\nabla M_0}{\sqrt{1+|\nabla M_0|^2}}.\label{coeff-M-eq-final-b-part}
	\end{align}
\end{subequations}
This is a genuinely macroscopic reaction-diffusion-taxis PDE for the leading term $M_0$ in the Hilbert expansion of the macroscopic glioma density $M$, thus it is supposed to approximate the tumor density dynamics for $\epsilon \to 0$.\footnote{This is just a formal deduction; a rigorous study of convergence raises considerable challenges and goes beyond the scope of this work.} The rest of equations in~\eqref{nondimensionalized_system} were already macroscopic.\\[-2ex]

For convenience of notation we will subsequently write $M$ instead of $M_0$. We summarize the full macroscopic system characterizing glioma dynamics under the influence of tissue, acidity, and vasculature:
\begin{subequations}\label{full-system-macro-nondim}
	\begin{align}
		&\partial_tM+\nabla_x\cdot\left(g(y^*)M\right)=\eta(h,s^*)\left(y^*-(y^*)^2\right)M(1-M),\label{full-system-macro-nondim-M}\\
		&\partial_tQ=c_1Q(1-Q-M)-c_2\frac{h}{1+h}Q,\label{full-system-macro-nondim-Q}\\
		&\partial_th=D_h\Delta h+\gamma (1-h)\frac{M}{1+M}-\delta he,\label{full-system-macro-nondim-h}\\
		&\partial_te=D_e\Delta e-\varsigma_e\nabla\cdot\left(e(1-e)\nabla h\right)+G_e(h,M)e(1-e),\label{full-system-macro-nondim-e}	
	\end{align}
\end{subequations}
with coefficients given in~\eqref{coeff-M-eq-final} and with $\eta (h,s^*)$  and $G_e(h,M)$ as in~\eqref{nondimensionalized_system-beta-eta} and~\eqref{nondimensionalized_system-G_e}, respectively. The system features self-diffusion, repellent pH-taxis, and haptotaxis, all of which involve limited fluxes. The diffusivity, tactic sensitivity functions, and even the proliferation rate depend on the solution components, directly or via the steady-state $y^*$ of receptor binding dynamics. Thus, although macroscopic, they still carry information from the lowermost (subcellular) level modeled here.\\[-2ex]

So far we considered the space variable $x\in \R^N$, however we should actually deal with a bounded region in which glioma cells, normal tissue, acidity, and endothelial cells are evolving. Let $\Om \subset \R^N$ be such bounded domain, with a smooth enough boundary. Through the rescaling $x\to \eps x$, the domain on which~\eqref{full-system-macro-nondim} holds is $\tilde \Om =\eps \Om$, with outer unit normal vector $\nu (x)$ at $x\in \partial \tilde \Om$. We are therefore interested in the boundary conditions on $\partial \tilde \Om$. Assuming no normal mass flux across the boundary gives the mesoscopic no-flux condition \cite{Plaza}
\begin{equation}\label{BC-plaza}
	\int _V\int _Yvp(t,x,v,y)\cdot \nu (x)\ dy\ dv=M^v(t,x)\cdot \nu (x)=0,\qquad \text{for all }x\in \partial \tilde \Om,\ t>0.
\end{equation}
Following \cite{Plaza} we write the boundary of the phase space as 
$$\partial \tilde \Om \times V\times Y=(\Gamma _+\cup \Gamma _-\cup \Gamma _0)\times Y,$$
where 
$$\Gamma _\pm :=\{(x,v)\in \partial \tilde \Om\times V\ :\ \pm v\cdot \nu (x)>0\},\quad \Gamma _0:=\{(x,v)\in \partial \tilde \Om\times V\ :\ v\cdot \nu (x)=0\}.$$

We assume that $\Gamma _0$ has zero measure w.r.t. the Lebesgue measure on $\partial \tilde \Om \times V$ and consider the trace spaces 
$$L^2_\pm:=L^2(\Gamma _\pm \times Y;\ |v\cdot \nu (x)|d\sigma (x)dvdz).$$

Moreover, $p$ is supposed to be regular enough so that we can define the traces $p|_{\Gamma _\pm \times Z}\in L^2_\pm$, and that for a fixed $t>0$ 
$$p|_{\partial \tilde \Om \times V\times Y}(t,x,v,y)=\lim _{\substack{\tilde x\in \tilde \Om \\
		\tilde x\to x}}p(t,\tilde x,y),\quad \text{for each }x\in \partial \tilde \Om.$$
Assuming that a regular Hilbert expansion is valid in $\tilde \Om $ we can therefore compute the trace by simply passing to the corresponding limit in the Hilbert expansions for $p(t,x,v,y)$ and accordingly also for the moments, in particular for $M^v$. 
Thus, the no-flux condition~\eqref{BC-plaza} becomes (at leading order, also recall our previous convention of using the notation $M$ for $M_0$):
\begin{equation}\label{BC-noflux-glioma}
	M^v(t,x)\cdot \nu (x)=g(y^*)M(t,x)\cdot \nu (x)=0, \quad x\in \partial \tilde \Om,\ t>0.
\end{equation}
upon using~\eqref{M0v}. The other PDEs in~\eqref{full-system-macro-nondim} were introduced in Subsection~\ref{subsec:acid-vasc} directly on a macroscopic level, thus we can simply impose no-flux conditions:
\begin{subequations}\label{BC-noflux-rest}
	\begin{align}
		D_h\nabla h\cdot \nu &=0\quad \text{on }\partial \tilde \Om,\ t>0,\\
		D_e\nabla e\cdot \nu &=0\quad \text{on }\partial \tilde \Om,\ t>0.
	\end{align}
\end{subequations}
To simplify notation we will use in the following $\Om $ instead of $\tilde \Om$.\\[-2ex]

System~\eqref{full-system-macro-nondim} with boundary conditions~\eqref{BC-noflux-glioma},~\eqref{BC-noflux-rest} has to be supplemented with adequate initial conditions. These can be the tumor
cell distribution (or an approximation of it) observed at diagnosis, some estimate of the macroscopic volume fraction of the tissue (e.g., most
simply FA, as in \cite{Corbin2018,EHKS} or assessed from DTI data as in \cite{ConteSurulescu,EHS,Hunt2016,kumar2020}), some (estimated) acidity distribution at diagnosis, and a given distribution of endothelial cell density.

\subsection{Invariant sets of regular solution components}\label{subsec:bound-M}
In this section we prove boundedness and nonnegativity of the components of a sufficiently smooth solution to (\ref{full-system-macro-nondim}). 
We first prove the following lemma:
\begin{lem}\label{degenerate-comparison-principle}
	Let $u\in\R$ and let $M\in C^{1,2}((0,T)\times{\Omega})$ be a classical solution to
	\begin{align}\label{equation_degenerate_maximum_principle}
		M_t&=\nabla \cdot (a(t,x,M,\nabla M)(u-M)\nabla M)+\nabla \cdot (b(t,x,M,\nabla M)(u-M))+c(t,x,M),\\
		0&=(a(t,x,M,\nabla M)(u-M)\nabla M+b(t,x,M,\nabla M)(u-M))\cdot \nu\mbox{\;\;on }\partial\Omega,\\
		M(0,x)&=M_0(x)\leq u,
	\end{align}
	where $a:(0,T)\times\Omega\times\R\times\R^n\rightarrow\R^{n\times n},\;b:(0,T)\times\Omega\times\R\times\R^n\rightarrow\R^n$ are continuously differentiable in all variables, and $c:(0,T)\times\Omega\times\R\rightarrow\R$ is continuous in all variables and Lipschitz w.r.t. $M$ on $[u-\epsilon,u+\epsilon]$ for some $\epsilon>0$. Let further $\xi^Ta(t,x,M,\nabla M)\xi\geq0$ and let $c(t,x,u)=0$. Then $M(t,x)\leq u$ for all $(t,x)\in(0,T)\times\Omega$.
\end{lem}

\proof
Assume there exist $(t_0,x_0)\in(0,T)\times\bar{\Omega}$, such that $M(t_0,x_0)$ is a (not necessarily strict) maximum of $M(t_0,\cdot)$ with $M(t_0,x_0)>u$. Consider now a $C^1$ path $z:[\tilde{t_0},t_0]\rightarrow \bar \Omega$ of (local) maxima of $M$ with $M(\tilde{t_0},z(\tilde{t_0}))<u$ and $z(t_0)=x_0$. As $M\in C^{1,2}((0,T)\times\Omega)$, such a path indeed exists. Define $Z(t):=M(t,z(t))$. Now we distinguish three cases:
\begin{itemize}
	\item[(i)] The point where $M$ intersects the value $u$ for the first time lies in the interior of $\Omega$. In this case, $(t_0,x_0)$ can be chosen such that $x_0\in\Omega$. Then the whole path $z(t)$ can be chosen to lie in the interior of $\Omega$ (after possibly shortening the time interval $[\tilde{t_0},t_0]$). Then, as $M$ has a maximum in $z(t)$ for each $t\in[\tilde{t_0},t_0],$ it holds $\nabla M(t,z(t))=0$. Now we find
	\begin{align*}
		\frac{\dd Z}{\dd t}=&\underbrace{\frac{\partial M}{\partial z}}_{=\nabla M=0}\frac{d z}{dt}+\frac{\partial M}{\partial t}=\frac{\partial M}{\partial t}\\
		=&\nabla\cdot\Big((u-M)a(t,z(t),M,\nabla M)\nabla M\Big)+\nabla\cdot\Big ((u-M)b(t,z(t),M,\nabla M)\Big )+c(t,z(t),M)\\
		=&-\nabla M \cdot \Big (a(t,z(t),M,\nabla M)\nabla M+b(t,z(t),M,\nabla M)\Big)\\
		&+(u-M)\nabla\cdot \Big (a(t,z(t),M,\nabla M)\nabla M+b(t,z(t),M,\nabla M)\Big )+c(t,z(t),M)\\
		=&(u-M)\nabla\cdot\Big (a(t,z(t),M,\nabla M)\nabla M+b(t,z(t),M,\nabla M)\Big )+c(t,z(t),M)\\
		=&(u-Z)\nabla\cdot \Big (a(t,z(t),M,\nabla M)\nabla M+b(t,z(t),M,\nabla M)\Big )+c(t,z(t),Z).
	\end{align*}
	\item[(ii)] The path lies completely on $\partial\Omega$ (after possibly shortening the time interval $[\tilde{t_0},t_0]$).\\
	By the boundary condition it holds $((u-M)a(t,x,M,\nabla M)\nabla M+(u-M)b(t,x,M,\nabla M))\cdot \nu=0$. Since $M(t,z(t))$ is a maximum on $\partial\Omega$, it holds $\nabla M\cdot \nu_{\bot}=0$ for all $\nu_{\bot}\bot \nu$ (otherwise, there would be an increase on $\partial\Omega$ and $M(t,z(t))$ could not be a maximum). Hence, we find
	\begin{align}\label{gradient_orthogonal_to_boundary_flux}
		\Big ((u-M)a(t,x,M,\nabla M)\nabla M+(u-M)b(t,x,M,\nabla M)\Big )\cdot\nabla M=0.
	\end{align}
	Furthermore, $\frac{dz}{dt}\bot \nu,$ since $z(t)$ lies by assumption completely on $\partial\Omega$. Hence, $\nabla M\cdot\frac{dz}{dt}=0$ and we find
	$$\frac{\dd Z}{\dd t}=\frac{\partial M}{\partial z}\frac{dz}{dt}+\frac{\partial M}{\partial t}=\frac{\partial M}{\partial t}.$$\\
	Now we have to distinguish again between two cases: 
	\begin{itemize}
		\item[(ii.a)] $M(t_1,z(t_1))=u$ for some $t_1\in(\tilde{t_0},t_0)$ and $M(t,z(t))\neq u$ in a neighbourhood of $t_1$: For $M(t,z(t))\neq u$, we divide (\ref{gradient_orthogonal_to_boundary_flux}) by $u-M$ to obtain
		$$\Big (a(t,z(t),M,\nabla M)\nabla M+b(t,z(t),M,\nabla M)\Big )\cdot\nabla M=0.$$ Since it holds $\Big (a(t,z(t),M,\nabla M)\nabla M+b(t,z(t),M,\nabla M)\Big )\cdot\nabla M=0$ everywhere except in $t_1$, by the continuity of all involved functions this also holds true in $t_1$.
		\item[(ii.b)] $M(t,z(t))=u$ on some closed time interval: Then on the boundary points $\tilde{t_0}$ and $t_0$ we can use the same argumentation as in the case above, to obtain $\Big (a(t,z(t),M,\nabla M)\nabla M+b(t,z(t),M,\nabla M)\Big )\cdot\nabla M=0$. In the interior of the interval, $M(t,z(t))=Z(t)$ is constantly $u$, hence it holds $\frac{\dd Z}{\dd t}=0$. Then, using $M=u$, we find
		\begin{align*}
			0=&\frac{\dd Z}{\dd t}=\frac{\partial M}{\partial t}\\
			=&\nabla\cdot\Big ((u-M)a(t,z(t),M,\nabla M)\nabla M\Big )+\nabla\cdot \Big ((u-M)b(t,z(t),M,\nabla M)\Big )+c(t,z(t),M)\\
			=&-\nabla M \cdot \Big (a(t,z(t),M,\nabla M)\nabla M+b(t,z(t),M,\nabla M)\Big ).\\
			&+(u-M)\nabla\cdot\Big (a(t,z(t),M,\nabla M)\nabla M+b(t,z(t),M,\nabla M)\Big )+c(t,z(t),M)\\
			=&-\nabla M \cdot \Big (a(t,z(t),M,\nabla M)\nabla M+b(t,z(t),M,\nabla M)\Big ).
		\end{align*}
	\end{itemize}
	Hence, in both cases we find $\nabla M \cdot \Big (a(t,z(t),M,\nabla M)\nabla M+b(t,z(t),M,\nabla M)\Big )=0$ and conclude
	\begin{align*}
		\frac{\dd Z}{\dd t}=&\frac{\partial M}{\partial t}\\
		=&\nabla\cdot\Big ((u-M)a(t,z(t),M,\nabla M)\nabla M\Big )+\nabla\cdot\Big ((u-M)b(t,z(t),M,\nabla M)\Big )+c(t,z(t),M)\\
		=&-\nabla M \cdot \Big(a(t,z(t),M,\nabla M)\nabla M+b(t,z(t),M,\nabla M)\Big )\\
		&+(u-M)\nabla\cdot\Big (a(t,z(t),M,\nabla M)\nabla M+b(t,z(t),M,\nabla M)\Big )+c(t,z(t),M)\\
		=&(u-M)\nabla\cdot\Big (a(t,z(t),M,\nabla M)\nabla M+b(t,z(t),M,\nabla M)\Big )+c(t,z(t),M)\\
		=&(u-Z)\nabla\cdot \Big (a(t,z(t),M,\nabla M)\nabla M+b(t,z(t),M,\nabla M)\Big )+c(t,z(t),Z).
	\end{align*}
	\item[(iii)] The path begins in the interior of $\Omega$ and intersects the value $u$ on $\partial\Omega$: In this case, the result $$\frac{\dd Z}{\dd t}=(u-Z)\nabla\cdot(a(t,z(t),M,\nabla M)\nabla M+b(t,z(t),M,\nabla M))+c(t,z(t),Z)$$ is obtained by combination of the cases above. 
\end{itemize}
We now interpret $\nabla\cdot \Big (a(t,z(t),M,\nabla M)\nabla M+b(t,z(t),M,\nabla M)\Big )$ on the path $z(t)$ as a function of time rather than a function of $M$, so $\nabla\cdot \Big (a(t,z(t),M,\nabla M)\nabla M+b(t,z(t),M,\nabla M)\Big )=:k(t)$. Then we obtain an ODE
$$\stackrel{\cdot}{Z}=k(t)(u-Z)+c(t,z(t),Z).$$ As the right hand side is Lipschitz continuous w.r.t. $Z$ on the interval $[u-\epsilon,u+\epsilon]$, there exists a unique solution to any initial value $Z(\tilde{t_0})$ in $[u-\epsilon,u+\epsilon]$. For initial value $u$, $Z\equiv u$ is the unique solution. For initial data in $[u-\epsilon,u)$, this solution cannot be intersected. Hence, $Z(t)\leq u$ for all $t\in[\tilde{t_0},t_0]$, which is a contradiction to $Z(t_0)>u$. This proves $M(t,x)\leq u$ for all $(t,x)\in(0,T)\times\Omega$.
\begin{remark}\label{rem:nach-lemma}
	Analogously, for $(u-M)$ replaced by $(M-u)$ in equation~\ref{equation_degenerate_maximum_principle} and initial data $M_0>u,$ one can prove $M(t,x)\geq u$ by defining a path of local minima instead of maxima.
\end{remark}
Now we are in a position to prove the following result:

\begin{lem}\label{bounded-solution}
	Let $(M,Q,h,e)\in \left(C^{1,2}((0,T)\times\Omega)\right)^4$, $T>0,$ be a classical solution to system (\ref{full-system-macro-nondim}) with boundary conditions (\ref{BC-noflux-glioma}) and (\ref{BC-noflux-rest}) and initial data $M_0(x),Q_0(x),h_0(x),e_0(x)\in[0,1]$ for all $x\in\Omega$. Then it holds $0\leq M(t,x),Q(t,x),h(t,x),e(t,x)\leq1$ for all $(t,x)\in(0,T)\times\Omega$.
\end{lem}
\proof

Applying Lemma~\ref{degenerate-comparison-principle} and its Remark~\ref{rem:nach-lemma} to equation (\ref{full-system-macro-nondim-M}), we find $0\leq M\leq 1$.
By application of a standard comparison principle for PDEs\footnote{\cite{DanersMedina}, Theorem 13.5} to (\ref{full-system-macro-nondim-h}), we find $0\leq h\leq 1$.

Bringing (\ref{full-system-macro-nondim-e}) into non-divergence form we can apply the same theorem to obtain $0\leq e\leq1$.

Finally, consider equation (\ref{full-system-macro-nondim-Q}). Obviously, $0$ is a subsolution, so $0\leq Q$. As we already showed nonnegativity of $h$, $1$ is a supersolution of (\ref{full-system-macro-nondim-Q}) and we conclude $Q\leq 1$.\\[-2ex]

\mysection{Numerical simulations}\label{sec:numerics}

\pgfplotsset{
	colormap={ECM}{
		rgb255=(255.000, 255.000, 255.000)
		rgb255=(247.714, 247.714, 247.714)
		rgb255=(240.429, 240.429, 240.429)
		rgb255=(233.143, 233.143, 233.143)
		rgb255=(225.857, 225.857, 225.857)
		rgb255=(218.571, 218.571, 218.571)
		rgb255=(211.286, 211.286, 211.286)
		rgb255=(204.000, 204.000, 204.000)
		rgb255=(200.821, 200.821, 200.821)
		rgb255=(197.641, 197.641, 197.641)
		rgb255=(194.462, 194.462, 194.462)
		rgb255=(191.282, 191.282, 191.282)
		rgb255=(188.103, 188.103, 188.103)
		rgb255=(184.923, 184.923, 184.923)
		rgb255=(181.744, 181.744, 181.744)
		rgb255=(178.564, 178.564, 178.564)
		rgb255=(175.385, 175.385, 175.385)
		rgb255=(172.205, 172.205, 172.205)
		rgb255=(169.026, 169.026, 169.026)
		rgb255=(165.846, 165.846, 165.846)
		rgb255=(162.667, 162.667, 162.667)
		rgb255=(159.487, 159.487, 159.487)
		rgb255=(156.308, 156.308, 156.308)
		rgb255=(153.128, 153.128, 153.128)
		rgb255=(149.949, 149.949, 149.949)
		rgb255=(146.769, 146.769, 146.769)
		rgb255=(143.590, 143.590, 143.590)
		rgb255=(140.410, 140.410, 140.410)
		rgb255=(137.231, 137.231, 137.231)
		rgb255=(134.051, 134.051, 134.051)
		rgb255=(130.872, 130.872, 130.872)
		rgb255=(127.692, 127.692, 127.692)
		rgb255=(124.513, 124.513, 124.513)
		rgb255=(121.333, 121.333, 121.333)
		rgb255=(118.154, 118.154, 118.154)
		rgb255=(114.974, 114.974, 114.974)
		rgb255=(111.795, 111.795, 111.795)
		rgb255=(108.615, 108.615, 108.615)
		rgb255=(105.436, 105.436, 105.436)
		rgb255=(102.256, 102.256, 102.256)
		rgb255=(99.077, 99.077, 99.077)
		rgb255=(95.897, 95.897, 95.897)
		rgb255=(92.718, 92.718, 92.718)
		rgb255=(89.538, 89.538, 89.538)
		rgb255=(86.359, 86.359, 86.359)
		rgb255=(83.179, 83.179, 83.179)
		rgb255=(80.000, 80.000, 80.000)
		rgb255=(75.294, 75.294, 75.294)
		rgb255=(70.588, 70.588, 70.588)
		rgb255=(65.882, 65.882, 65.882)
		rgb255=(61.176, 61.176, 61.176)
		rgb255=(56.471, 56.471, 56.471)
		rgb255=(51.765, 51.765, 51.765)
		rgb255=(47.059, 47.059, 47.059)
		rgb255=(42.353, 42.353, 42.353)
		rgb255=(37.647, 37.647, 37.647)
		rgb255=(32.941, 32.941, 32.941)
		rgb255=(28.235, 28.235, 28.235)
		rgb255=(23.529, 23.529, 23.529)
		rgb255=(18.824, 18.824, 18.824)
		rgb255=(14.118, 14.118, 14.118)
		rgb255=(9.412, 9.412, 9.412)
		rgb255=(4.706, 4.706, 4.706)
		rgb255=(0.000, 0.000, 0.000)
	},
	colormap={glioma}{
		rgb255=(255.000, 245.000, 235.000)
		rgb255=(254.997, 243.094, 232.067)
		rgb255=(254.937, 241.233, 228.926)
		rgb255=(254.832, 239.395, 225.572)
		rgb255=(254.693, 237.556, 222.000)
		rgb255=(254.529, 235.694, 218.205)
		rgb255=(254.350, 233.785, 214.180)
		rgb255=(254.163, 231.807, 209.922)
		rgb255=(253.977, 229.735, 205.425)
		rgb255=(253.797, 227.548, 200.684)
		rgb255=(253.628, 225.222, 195.696)
		rgb255=(253.475, 222.735, 190.455)
		rgb255=(253.340, 220.063, 184.959)
		rgb255=(253.224, 217.183, 179.202)
		rgb255=(253.128, 214.072, 173.181)
		rgb255=(253.049, 210.706, 166.893)
		rgb255=(252.985, 207.062, 160.335)
		rgb255=(252.932, 203.140, 153.528)
		rgb255=(252.885, 198.992, 146.537)
		rgb255=(252.845, 194.671, 139.432)
		rgb255=(252.817, 190.233, 132.282)
		rgb255=(252.809, 185.732, 125.152)
		rgb255=(252.836, 181.219, 118.107)
		rgb255=(252.915, 176.749, 111.209)
		rgb255=(253.067, 172.373, 104.519)
		rgb255=(253.286, 168.109, 98.063)
		rgb255=(253.525, 163.927, 91.816)
		rgb255=(253.737, 159.795, 85.752)
		rgb255=(253.870, 155.682, 79.844)
		rgb255=(253.876, 151.557, 74.064)
		rgb255=(253.703, 147.391, 68.387)
		rgb255=(253.302, 143.154, 62.783)
		rgb255=(252.623, 138.819, 57.225)
		rgb255=(251.658, 134.383, 51.706)
		rgb255=(250.439, 129.865, 46.249)
		rgb255=(249.002, 125.286, 40.875)
		rgb255=(247.380, 120.665, 35.605)
		rgb255=(245.608, 116.020, 30.466)
		rgb255=(243.720, 111.370, 25.489)
		rgb255=(241.752, 106.733, 20.720)
		rgb255=(239.732, 102.124, 16.232)
		rgb255=(237.621, 97.561, 12.060)
		rgb255=(235.332, 93.064, 8.300)
		rgb255=(232.779, 88.662, 5.491)
		rgb255=(229.875, 84.391, 3.534)
		rgb255=(226.536, 80.296, 2.245)
		rgb255=(222.678, 76.428, 1.465)
		rgb255=(218.218, 72.845, 1.058)
		rgb255=(213.092, 69.599, 0.909)
		rgb255=(207.358, 66.695, 0.949)
		rgb255=(201.138, 64.102, 1.130)
		rgb255=(194.551, 61.777, 1.413)
		rgb255=(187.716, 59.675, 1.763)
		rgb255=(180.746, 57.746, 2.152)
		rgb255=(173.755, 55.940, 2.555)
		rgb255=(166.851, 54.213, 2.952)
		rgb255=(160.142, 52.522, 3.325)
		rgb255=(153.730, 50.833, 3.662)
		rgb255=(147.716, 49.115, 3.951)
		rgb255=(142.200, 47.337, 4.183)
		rgb255=(137.276, 45.473, 4.338)
		rgb255=(133.040, 43.491, 4.357)
		rgb255=(129.585, 41.352, 4.241)
		rgb255=(127.000, 39.000, 4.000)
	},
	colormap={AC}{
		rgb255=(0.000, 68.000, 27.000)
		rgb255=(0.000, 74.369, 28.400)
		rgb255=(0.000, 80.353, 30.025)
		rgb255=(0.000, 85.971, 31.886)
		rgb255=(0.000, 91.247, 33.980)
		rgb255=(0.000, 96.208, 36.295)
		rgb255=(0.000, 100.885, 38.813)
		rgb255=(0.000, 105.309, 41.511)
		rgb255=(0.486, 109.515, 44.364)
		rgb255=(4.950, 113.540, 47.346)
		rgb255=(10.399, 117.421, 50.431)
		rgb255=(15.766, 121.197, 53.594)
		rgb255=(20.520, 124.911, 56.813)
		rgb255=(24.842, 128.602, 60.063)
		rgb255=(28.810, 132.316, 63.326)
		rgb255=(32.461, 136.095, 66.578)
		rgb255=(35.809, 139.985, 69.802)
		rgb255=(38.924, 144.005, 72.984)
		rgb255=(41.988, 148.120, 76.124)
		rgb255=(45.170, 152.288, 79.222)
		rgb255=(48.615, 156.467, 82.278)
		rgb255=(52.447, 160.612, 85.292)
		rgb255=(56.764, 164.679, 88.264)
		rgb255=(61.643, 168.621, 91.192)
		rgb255=(67.130, 172.391, 94.077)
		rgb255=(73.161, 175.969, 96.935)
		rgb255=(79.558, 179.374, 99.811)
		rgb255=(86.174, 182.626, 102.750)
		rgb255=(92.892, 185.748, 105.799)
		rgb255=(99.618, 188.764, 109.003)
		rgb255=(106.277, 191.698, 112.410)
		rgb255=(112.804, 194.575, 116.065)
		rgb255=(119.144, 197.420, 120.013)
		rgb255=(125.285, 200.243, 124.253)
		rgb255=(131.251, 203.033, 128.738)
		rgb255=(137.062, 205.778, 133.419)
		rgb255=(142.733, 208.469, 138.249)
		rgb255=(148.278, 211.095, 143.178)
		rgb255=(153.707, 213.643, 148.158)
		rgb255=(159.030, 216.103, 153.141)
		rgb255=(164.254, 218.463, 158.080)
		rgb255=(169.380, 220.724, 162.955)
		rgb255=(174.405, 222.892, 167.767)
		rgb255=(179.326, 224.972, 172.515)
		rgb255=(184.140, 226.971, 177.201)
		rgb255=(188.842, 228.897, 181.824)
		rgb255=(193.431, 230.757, 186.385)
		rgb255=(197.901, 232.558, 190.885)
		rgb255=(202.250, 234.306, 195.322)
		rgb255=(206.471, 236.002, 199.688)
		rgb255=(210.555, 237.641, 203.968)
		rgb255=(214.493, 239.218, 208.146)
		rgb255=(218.277, 240.731, 212.210)
		rgb255=(221.896, 242.176, 216.143)
		rgb255=(225.342, 243.548, 219.931)
		rgb255=(228.605, 244.844, 223.559)
		rgb255=(231.677, 246.059, 227.011)
		rgb255=(234.547, 247.190, 230.273)
		rgb255=(237.206, 248.234, 233.330)
		rgb255=(239.645, 249.186, 236.166)
		rgb255=(241.853, 250.042, 238.766)
		rgb255=(243.822, 250.799, 241.115)
		rgb255=(245.540, 251.453, 243.198)
		rgb255=(247.000, 252.000, 245.000)
	},
	colormap={endothelial}{
		rgb255=(252.000, 251.000, 253.000)
		rgb255=(250.640, 249.285, 252.017)
		rgb255=(249.212, 247.577, 251.033)
		rgb255=(247.709, 245.864, 250.044)
		rgb255=(246.121, 244.131, 249.046)
		rgb255=(244.439, 242.366, 248.033)
		rgb255=(242.655, 240.557, 247.001)
		rgb255=(240.758, 238.690, 245.946)
		rgb255=(238.741, 236.754, 244.863)
		rgb255=(236.593, 234.734, 243.747)
		rgb255=(234.307, 232.619, 242.593)
		rgb255=(231.872, 230.397, 241.398)
		rgb255=(229.280, 228.055, 240.155)
		rgb255=(226.521, 225.580, 238.862)
		rgb255=(223.586, 222.963, 237.512)
		rgb255=(220.467, 220.189, 236.101)
		rgb255=(217.153, 217.249, 234.624)
		rgb255=(213.649, 214.133, 233.075)
		rgb255=(209.985, 210.841, 231.438)
		rgb255=(206.194, 207.372, 229.702)
		rgb255=(202.311, 203.724, 227.853)
		rgb255=(198.370, 199.898, 225.878)
		rgb255=(194.404, 195.893, 223.764)
		rgb255=(190.450, 191.708, 221.499)
		rgb255=(186.542, 187.342, 219.070)
		rgb255=(182.692, 182.822, 216.493)
		rgb255=(178.888, 178.210, 213.824)
		rgb255=(175.112, 173.568, 211.119)
		rgb255=(171.346, 168.961, 208.435)
		rgb255=(167.574, 164.450, 205.826)
		rgb255=(163.778, 160.098, 203.349)
		rgb255=(159.941, 155.965, 201.059)
		rgb255=(156.042, 152.109, 199.006)
		rgb255=(152.090, 148.508, 197.180)
		rgb255=(148.115, 145.064, 195.508)
		rgb255=(144.155, 141.674, 193.915)
		rgb255=(140.247, 138.239, 192.326)
		rgb255=(136.433, 134.659, 190.665)
		rgb255=(132.754, 130.835, 188.858)
		rgb255=(129.256, 126.670, 186.832)
		rgb255=(125.978, 122.072, 184.517)
		rgb255=(122.912, 117.050, 181.920)
		rgb255=(120.017, 111.685, 179.101)
		rgb255=(117.249, 106.057, 176.118)
		rgb255=(114.567, 100.244, 173.032)
		rgb255=(111.932, 94.323, 169.902)
		rgb255=(109.308, 88.370, 166.786)
		rgb255=(106.666, 82.461, 163.743)
		rgb255=(103.983, 76.669, 160.824)
		rgb255=(101.255, 71.015, 158.039)
		rgb255=(98.489, 65.494, 155.370)
		rgb255=(95.693, 60.101, 152.805)
		rgb255=(92.873, 54.827, 150.328)
		rgb255=(90.038, 49.664, 147.925)
		rgb255=(87.195, 44.599, 145.581)
		rgb255=(84.355, 39.618, 143.285)
		rgb255=(81.525, 34.699, 141.021)
		rgb255=(78.716, 29.816, 138.776)
		rgb255=(75.940, 24.926, 136.538)
		rgb255=(73.209, 19.963, 134.294)
		rgb255=(70.535, 14.815, 132.030)
		rgb255=(67.932, 9.261, 129.735)
		rgb255=(65.415, 4.137, 127.396)
		rgb255=(63.000, 0.000, 125.000)
	}
}
\graphicspath{{./}}



\begin{figure}[t] 
	\centering
	\figurewidth=\linewidth
	\def \expSetup {initial}
	\begin{tikzpicture}
		\begin{groupplot}[
			/tikz/mark size=1.5pt,
			group style={
				group name=my plots,
				group size=3 by 2,
				horizontal sep=1.6cm,      
				vertical sep=1.2cm,        
			},
			xmin=-2,
			xmax=2,
			ymin=-2,
			ymax=2,
			ylabel shift = 2 em,
			xtick = \empty,
			ytick= \empty,
			ticklabel style = {font=\scriptsize},
			axis line style = thick,
			axis background/.style={fill=white},
			width=.3\figurewidth,
			height=.3\figurewidth, 
			]
			\nextgroupplot[
			title = {glioma cells ($M$)},
			point meta min=0,
			point meta max=0.25,
			colormap name=glioma,
			colorbar,
			colorbar style={at={(1.1,1)},anchor=north west}
			]
			\addplot [forget plot] graphics [xmin=-2, xmax=2, ymin=-2, ymax=2] {\expSetup1};
			\nextgroupplot[
			title = {tissue ($Q$)},
			point meta min=0.8,
			point meta max=1,
			colormap name=ECM,
			colorbar,
			colorbar style={at={(1.1,1)},anchor=north west}
			]
			\addplot [forget plot] graphics [xmin=-2, xmax=2, ymin=-2, ymax=2] {\expSetup2};
			
			\nextgroupplot[
			title = {acidity (pH)},
			point meta min=6.5,
			point meta max=7.05,
			colormap name=AC,
			colorbar,
			colorbar style={at={(1.1,1)},anchor=north west}
			]
			\addplot [forget plot] graphics [xmin=-2, xmax=2, ymin=-2, ymax=2] {\expSetup3};
			\nextgroupplot[
			title = {endothelial cells ($e$)},
			point meta min=0,
			point meta max=0.9,
			colormap name=endothelial,
			colorbar,
			colorbar style={at={(1.1,1)},anchor=north west}
			]
			\addplot [forget plot] graphics [xmin=-2, xmax=2, ymin=-2, ymax=2] {\expSetup4};
			\nextgroupplot[
			title = {fractional anisotropy},
			point meta min=0,
			point meta max=0.9,
			colormap name=ECM,
			colorbar,
			colorbar style={at={(1.1,1)},anchor=north west}
			]
			\addplot [forget plot] graphics [xmin=-2, xmax=2, ymin=-2, ymax=2] {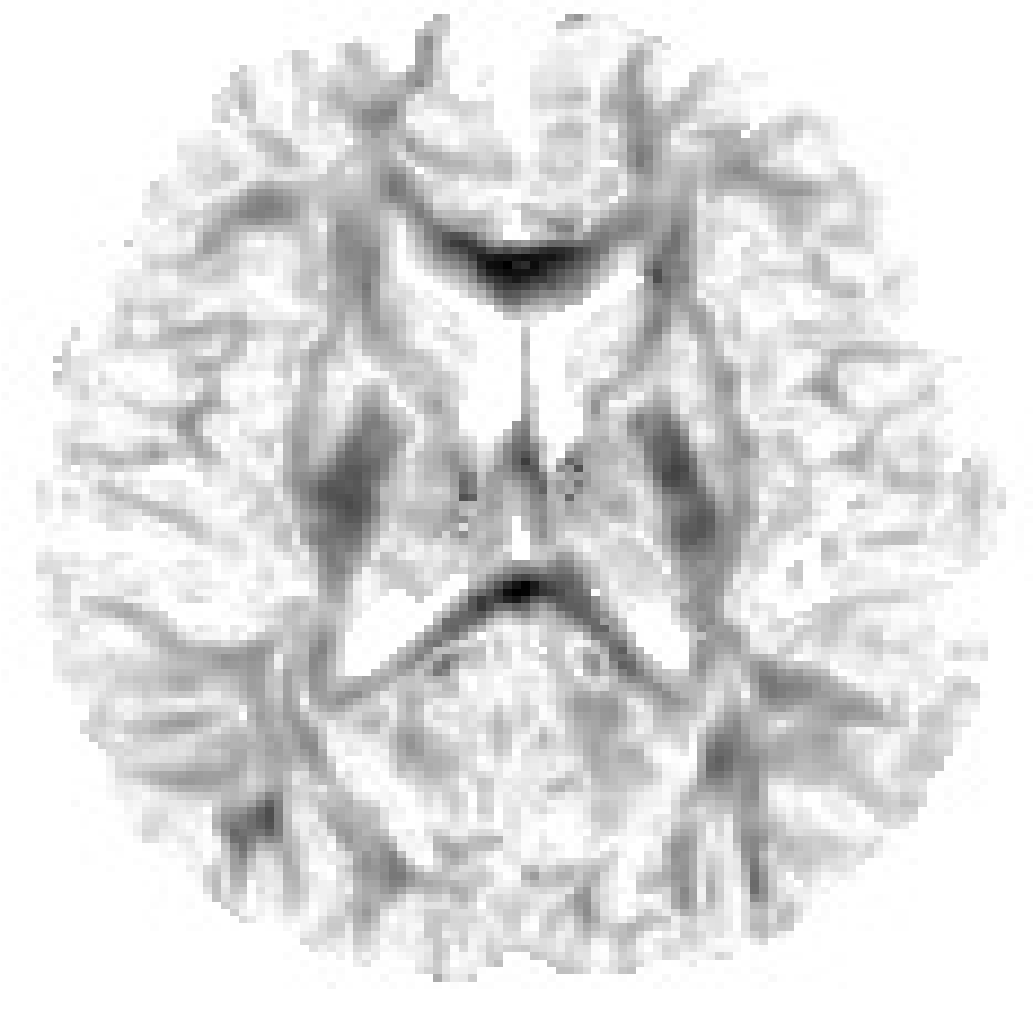};
		\end{groupplot}
	\end{tikzpicture}%
	\caption{Initial {\cb amounts} of the four {\cb solution components} $\left(M,Q,h,e\right)^T$ of the system~\eqref{full-system-macro-nondim} and fractional anisotropy on the spatial domain $[0,1]\times[0,1.2155]$. The {\cb forms} of the initial amounts are given in~\eqref{eq:ICs}.}\label{fig:initial}
\end{figure}
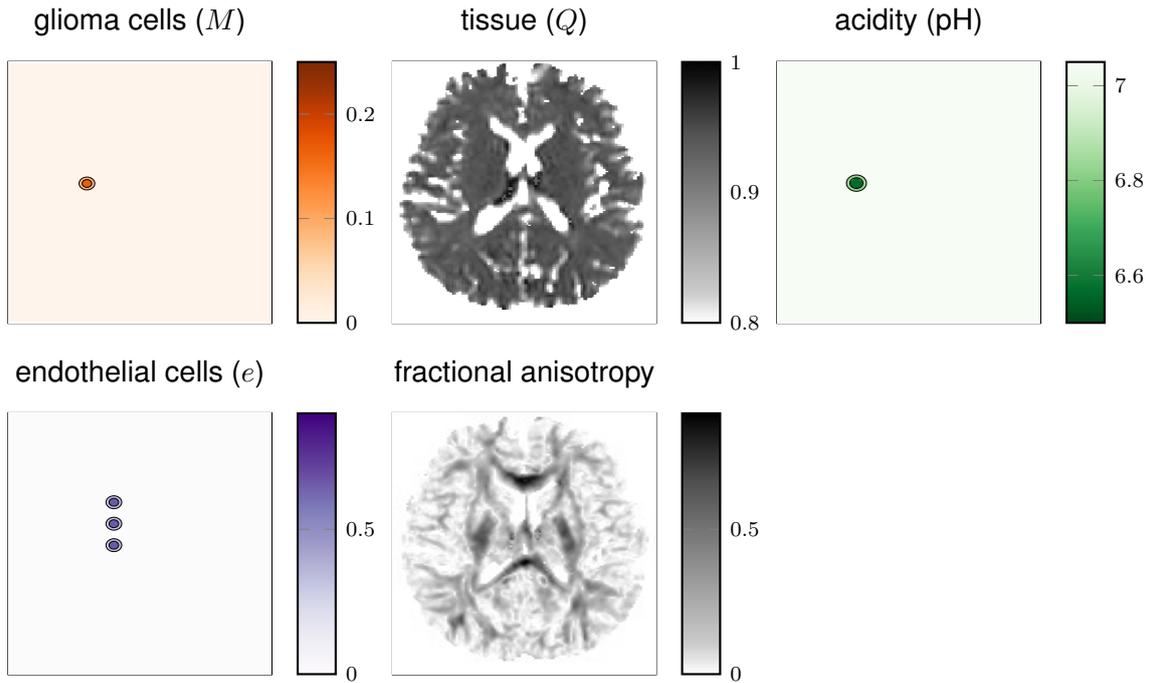

With a number of numerical experiments, we study the model~\eqref{full-system-macro-nondim} in its derived form and with slight modifications. To this end we employ a second order Finite Volume scheme on an equidistant mesh over an either rectangular domain in 2D ($200\times200$ control volumes) or a cuboid domain in 3D ($128\times128\times128$ control volumes) with no-flux conditions at the boundaries. The scheme employs central upwind fluxes obtained by discretizing~\eqref{eq:g} at the center of the mesh cell interfaces through central differences, averaging, and interpolation of the brain data. To prevent oscillatory behavior of solutions we use the \textit{minimized-central} slope limiter, \cite{van_leer_towards_1979}. For the time stepping we use the implicit-explicit Midpoint scheme from \cite{pareschi_implicitexplicit_2005}, which lets us treat the stiff diffusion of the acidity and of the endothelial cells implicitly. This strategy together with the limited fluxes in the model allow for large time increments in the computations. For more details on the method we refer to \cite{kolbe2020modeling, kolbe_study_2016, nn.2016}, where the same numerical approach was applied to similar 2D advection-reaction-diffusion problems, and to \cite{nam.2020, lnm.2020} where it was applied on a 2D and 3D hybrid atomistic-macroscopic cancer invasion model. The algorithms were implemented in MATLAB \cite{MATLAB.2019} and the visualisations were produced by MATLAB \cite{MATLAB.2019} and PARAVIEW \cite{paraview}.\\[-2ex]

The first numerical experiments that we consider (Experiments 1--4) are conducted over the spatial domain $\Omega_1 = [0,1]\times[0,1.2155]$ and over the time frame $t\in[0,25]$; the initial conditions are accordingly given, for every $(x,y)\in\Omega_1$, through
\begin{subequations}\label{eq:ICs}
	\begin{align}
		M_0(x,y) &=  0.25 e^{-\frac1\varepsilon \left( (x-0.3)^2+(y-0.65)^2 \right) },\\
		e_0(x,y) & =  e^{-\frac1\varepsilon \left( (x-0.4)^2+(y-0.8)^2 \right) } + 		e^{-\frac1\varepsilon \left( (x-0.4)^2+(y-0.7)^2\right) }+ e^{ -\frac1\varepsilon\left( (x-0.4)^2+(y-0.6)^2 \right) },\\
		h_0(x,y) &= 0.03M_0(x,y) + 10^{-2.8},
	\end{align}
\end{subequations}
where $\varepsilon=8\times 10^{-4}$. The initial condition $Q_0$ {\cb for the spatial distribution of brain tissue density} is given in~\eqref{tissue_ini}. Figure~\ref{fig:initial} shows the initial {\cb amounts (volume fractions)} of the four unknowns $\left(M,Q,h,e\right)^T$ of the system~\eqref{full-system-macro-nondim}. {\cb In this and all subsequent plots we convert the proton concentration $h$ into $pH$-values by $pH=-\log_{10}(h)$ and represent acidity by way of those values.}

\begin{figure}[t]
	\figurewidth=\linewidth
	\def \expSetup {reference_CNN}
	\begin{tikzpicture}
		\begin{groupplot}[
			/tikz/mark size=1.5pt,
			group style={
				group name=my plots,
				group size=4 by 3,
				horizontal sep=1cm,      
				vertical sep=0.7cm,        
			},
			xmin=-2,
			xmax=2,
			ymin=-2,
			ymax=2,
			ylabel shift = 2 em,
			xtick = \empty,
			ytick= \empty,
			ticklabel style = {font=\scriptsize},
			axis line style = thick,
			axis background/.style={fill=white},
			width=.25\figurewidth,
			height=.25\figurewidth, 
			]
			\nextgroupplot[ylabel = glioma cells ($M$), title = {$t=6.25$}]
			\addplot [forget plot] graphics [xmin=-2, xmax=2, ymin=-2, ymax=2] {\expSetup1};
			\nextgroupplot[title = {$t=12.5$}]
			\addplot [forget plot] graphics [xmin=-2, xmax=2, ymin=-2, ymax=2] {\expSetup2};
			\nextgroupplot[title = {$t=18.75$}]
			\addplot [forget plot] graphics [xmin=-2, xmax=2, ymin=-2, ymax=2] {\expSetup3};
			\nextgroupplot[
			title = {$t=25$},
			point meta min=0,
			point meta max=0.03,
			colormap name=glioma,
			colorbar,
			colorbar style={at={(1.2,1)},anchor=north west}
			]
			\addplot [forget plot] graphics [xmin=-2, xmax=2, ymin=-2, ymax=2] {\expSetup4};
			\nextgroupplot[ylabel = acidity ($pH$)]
			\addplot [forget plot] graphics [xmin=-2, xmax=2, ymin=-2, ymax=2] {\expSetup9};
			\nextgroupplot
			\addplot [forget plot] graphics [xmin=-2, xmax=2, ymin=-2, ymax=2] {\expSetup10};
			\nextgroupplot
			\addplot [forget plot] graphics [xmin=-2, xmax=2, ymin=-2, ymax=2] {\expSetup11};
			\nextgroupplot[
			point meta min=6.5,
			point meta max=7.05,
			colormap name=AC,
			colorbar,
			colorbar style={at={(1.2,1)},anchor=north west}
			]
			\addplot [forget plot] graphics [xmin=-2, xmax=2, ymin=-2, ymax=2] {\expSetup12};
			\nextgroupplot[ylabel = endothelial cells ($e$)]
			\addplot [forget plot] graphics [xmin=-2, xmax=2, ymin=-2, ymax=2] {\expSetup13};
			\nextgroupplot
			\addplot [forget plot] graphics [xmin=-2, xmax=2, ymin=-2, ymax=2] {\expSetup14};
			\nextgroupplot
			\addplot [forget plot] graphics [xmin=-2, xmax=2, ymin=-2, ymax=2] {\expSetup15};
			\nextgroupplot[
			point meta min=0,
			point meta max=0.8,
			colormap name=endothelial,
			colorbar,
			colorbar style={at={(1.2,1)},anchor=north west}
			]
			\addplot [forget plot] graphics [xmin=-2, xmax=2, ymin=-2, ymax=2] {\expSetup16};
		\end{groupplot}
	\end{tikzpicture}%
	
	\caption{Simulation results {\cb for} \textbf{Experiment~\ref{exp:dom-hap} --- dominant haptotaxis}. Time evolution (vertical columns) of {\cb glioma cell density} $M$, acidity $pH$, and endothelial cell density $e$ over the domain $[0,1]\times[0,1.2155]$. The glioma cells respond via haptotaxis to the anisotropic brain tissue. The acid, produced by the tumor cells, diffuses in the environment and serves as chemoattractant for the endothelial cells, and as degradation agent for the brain tissue, cf. Figure~\ref{fig:tissue}. The vascularization is more pronounced and directed towards {\cb lower} $pH$ levels {\cb (hence towards the main tumor mass)}.}\label{fig:exp1}	
\end{figure}
\begin{table}
	\centering
	\begin{tabular}{c|l|l}
		\hline
		Symbol&Description&Value\\
		\hline
		$D_h$	& acid diffusion& $10^{-4}$ \\
		$D_e$  & endothelial cell diffusion& $10^{-6}$\\
		$c_1$ & tissue proliferation& $3\times10^{-4}$ \\
		$\gamma$ & glioma production of acid & $10^{-2}$\\
		$\nu_e$ &endothelial cell proliferation& $5\times10^{-3}$\\
		$c_2$&acid degradation of tissue& $5\times10^{-3}$\\
		$\delta$&acid uptake by endothelial cells&$8\times10^{-4}$\\
		$\varsigma_e$&acidotaxis of endothelial cells &$1.5\times10^{-1}$\\ 
		$\rho_1$& weight of haptotaxis in glioma migration&$7.5\times10^{-1}$\\
		$\rho_2$&weight of diffusion in glioma migration&$1.5\times10^{-2}$\\
		$k$&tissue carrying capacity&$10^{-2}$\\
		$a_1, a_2$&glioma migration scaling&$1$\\
		\hline
	\end{tabular}
	\caption{Dimensionless parameters employed in \textbf{Experiment~\ref{exp:dom-hap} --- dominant haptotaxis} and in \textbf{Experiment~\ref{exp:3d} --- dominant haptotaxis in 3D}.}\label{tbl:ref_pars}
\end{table}

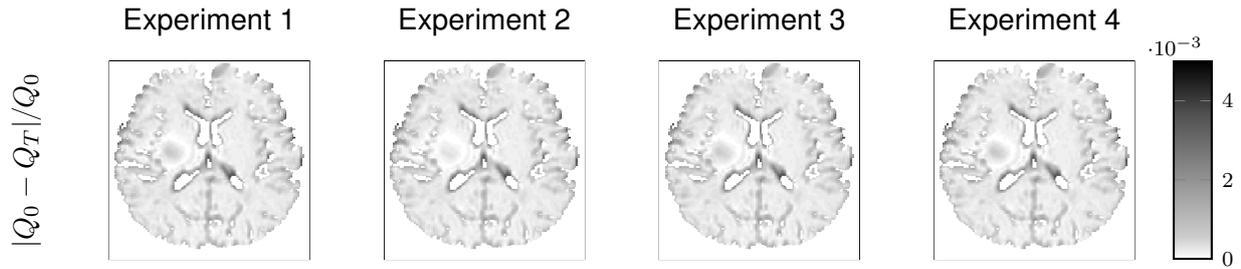
\begin{figure}[t] 
	\figurewidth=\linewidth
	\def \expSetup {tissue_diff}
	\begin{tikzpicture}
		\begin{groupplot}[
			/tikz/mark size=1.5pt,
			group style={
				group name=my plots,
				group size=4 by 1,
				horizontal sep=1cm,      
				vertical sep=0.7cm,        
			},
			xmin=-2,
			xmax=2,
			ymin=-2,
			ymax=2,
			ylabel shift = 2 em,
			xtick = \empty,
			ytick= \empty,
			ticklabel style = {font=\scriptsize},
			axis line style = thick,
			axis background/.style={fill=white},
			width=.25\figurewidth,
			height=.25\figurewidth, 
			]
			\nextgroupplot[ylabel = $|Q_0-Q_T|/Q_0$, title = {Experiment 1}]
			\addplot [forget plot] graphics [xmin=-2, xmax=2, ymin=-2, ymax=2] {\expSetup5};
			\nextgroupplot[title = {Experiment 2}]
			\addplot [forget plot] graphics [xmin=-2, xmax=2, ymin=-2, ymax=2] {\expSetup6};
			\nextgroupplot[title = {Experiment 3}]
			\addplot [forget plot] graphics [xmin=-2, xmax=2, ymin=-2, ymax=2] {\expSetup7};
			\nextgroupplot[
			title = {Experiment 4},
			point meta min=0,
			point meta max=0.005,
			colormap name=ECM,
			colorbar,
			colorbar style={at={(1.2,1)},anchor=north west}
			]
			\addplot [forget plot] graphics [xmin=-2, xmax=2, ymin=-2, ymax=2] {\expSetup8};
			
		\end{groupplot}
	\end{tikzpicture}%
	\caption{Relative difference between the initial ($Q_0$) and the final tissue density ($Q_T$) for all experiments studied here. Both the effects of tissue regeneration and degradation are visible. The tumor-related tissue degradation, in particular, is evident by the shadow cast on the {\cb acidic} region {\cb (due to hypoxia)}.}\label{fig:tissue}	
\end{figure}

\begin{experiment}\textbf{dominant haptotaxis.}\label{exp:dom-hap}
	In this first experiment we investigate the dynamics {\cb exhibited by} model~\eqref{full-system-macro-nondim} when augmented with the initial conditions~\eqref{eq:ICs} and {\cb using} the parameter set given in Table~\ref{tbl:ref_pars}. {\cb A particular feature} of this experiment is that glioma cell migration is dominated by haptotaxis rather than by random movement or negative acidotaxis, {\cb according to} the values of the respective weight parameters $\rho_1=0.75$,  $\rho_2=0.015$,  and $1-\rho_1-\rho_2=0.235$ in Table~\ref{tbl:ref_pars} along with their role in ~\eqref{coeff-M-eq-final-b-part}.\\[-2ex]
	
	The time evolution of {\cb numerically computed amounts of glioma cells $M$, acidity $pH$, and endothelial cells $e$ is} exhibited in Figure~\ref{fig:exp1} with the corresponding initial conditions shown in Figure~\ref{fig:initial}. {\cb The  glioma cells (of density $M$) respond to gradients of the (anisotropic) brain tissue (of density $Q$), while at the same time the tumor acts as source of protons (of concentration $h$). The acid, in turn, diffuses in the environment and serves as chemoattractant for the endothelial cells (of density $e$).} This justifies the more pronounced vascularization, directed towards {\cb lower} $pH$ levels. The acid {\cb (by way of hypoxia)} is also responsible for the degradation of brain tissue; this, along with the physiological regeneration of {\cb the extracellular matrix}, is visualized in the first  panel of Figure~\ref{fig:tissue} through a (relative) comparison between {\cb the tissue densities $Q_0$ and $Q_T$ at the initial and final computation times, respectively.}

\end{experiment}

\begin{figure}[t] 
	\figurewidth=\linewidth
	\def \expSetup {reference_CNN_acid}
	\begin{tikzpicture}
		\begin{groupplot}[
			/tikz/mark size=1.5pt,
			group style={
				group name=my plots,
				group size=4 by 3,
				horizontal sep=1cm,      
				vertical sep=0.7cm,        
			},
			xmin=-2,
			xmax=2,
			ymin=-2,
			ymax=2,
			ylabel shift = 2 em,
			xtick = \empty,
			ytick= \empty,
			ticklabel style = {font=\scriptsize},
			axis line style = thick,
			axis background/.style={fill=white},
			width=.25\figurewidth,
			height=.25\figurewidth, 
			]
			\nextgroupplot[ylabel = glioma cells ($M$), title = {$t=6.25$}]
			\addplot [forget plot] graphics [xmin=-2, xmax=2, ymin=-2, ymax=2] {\expSetup1};
			\nextgroupplot[title = {$t=12.5$}]
			\addplot [forget plot] graphics [xmin=-2, xmax=2, ymin=-2, ymax=2] {\expSetup2};
			\nextgroupplot[title = {$t=18.75$}]
			\addplot [forget plot] graphics [xmin=-2, xmax=2, ymin=-2, ymax=2] {\expSetup3};
			\nextgroupplot[
			title = {$t=25$},
			point meta min=0,
			point meta max=0.025,
			colormap name=glioma,
			colorbar,
			colorbar style={at={(1.2,1)},anchor=north west}
			]
			\addplot [forget plot] graphics [xmin=-2, xmax=2, ymin=-2, ymax=2] {\expSetup4};
			\nextgroupplot[ylabel = acidity (pH)]
			\addplot [forget plot] graphics [xmin=-2, xmax=2, ymin=-2, ymax=2] {\expSetup9};
			\nextgroupplot
			\addplot [forget plot] graphics [xmin=-2, xmax=2, ymin=-2, ymax=2] {\expSetup10};
			\nextgroupplot
			\addplot [forget plot] graphics [xmin=-2, xmax=2, ymin=-2, ymax=2] {\expSetup11};
			\nextgroupplot[
			point meta min=6.5,
			point meta max=7.05,
			colormap name=AC,
			colorbar,
			colorbar style={at={(1.2,1)},anchor=north west}
			]
			\addplot [forget plot] graphics [xmin=-2, xmax=2, ymin=-2, ymax=2] {\expSetup12};
			\nextgroupplot[ylabel = endothelial cells ($e$)]
			\addplot [forget plot] graphics [xmin=-2, xmax=2, ymin=-2, ymax=2] {\expSetup13};
			\nextgroupplot
			\addplot [forget plot] graphics [xmin=-2, xmax=2, ymin=-2, ymax=2] {\expSetup14};
			\nextgroupplot
			\addplot [forget plot] graphics [xmin=-2, xmax=2, ymin=-2, ymax=2] {\expSetup15};
			\nextgroupplot[
			point meta min=0,
			point meta max=0.8,
			colormap name=endothelial,
			colorbar,
			colorbar style={at={(1.2,1)},anchor=north west}
			]
			\addplot [forget plot] graphics [xmin=-2, xmax=2, ymin=-2, ymax=2] {\expSetup16};
		\end{groupplot}
	\end{tikzpicture}%
	
	\caption{Simulation results of \textbf{Experiment~\ref{exp:dom-ac} --- dominant acidotaxis.} In the same setting as in Experiment~\ref{exp:dom-hap}, the weight parameters $\rho_1$ and $\rho_2$ {\cb controlling the migration of glioma cells} have been set in favor of the {\cb repellent} acidotaxis. The {\cb main effect}, when compared to Figure~\ref{fig:exp1}, is the drop of glioma density {\cb at  the location of the initial tumor}, and the wider spread away from it.}\label{fig:exp2}	
\end{figure}
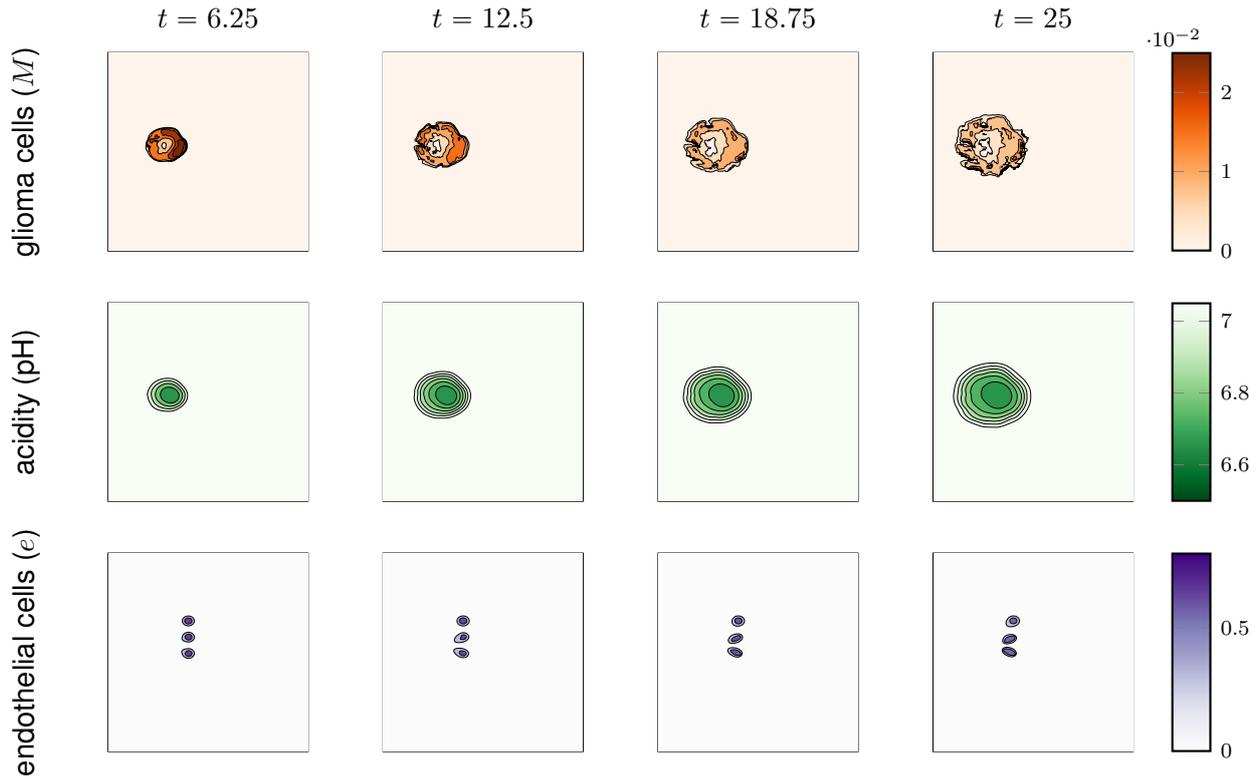

\begin{experiment}\textbf{dominant acidotaxis.}\label{exp:dom-ac}
	In this experiment we consider the same modeling setting as in Experiment~\ref{exp:dom-hap} augmented with the same initial conditions~\eqref{eq:ICs}, and the same parameter set given in Table~\ref{tbl:ref_pars}, except for the parameters $\rho_1$, $\rho_2$ {\cb weighting the motility behavior of glioma} cells. In particular, we consider in this experiment a glioma {\cb migration regime dominated by acidotaxis {\cb (meaning that the tumor cells are repelled by low pH)} and accordingly choose} $\rho_1=0.4$,  $\rho_2=0.015$,  and $1-\rho_1-\rho_2=0.585$.\\[-2ex]
	
	The time evolution of $(M, pH, e)$ is shown in Figure~\ref{fig:exp2}, which, similarly to the previous Experiment~\ref{exp:dom-hap}, exhibits {\cb the spread of glioma in the anisotropic brain tissue and a pronounced vascularization  towards the lower $pH$ region. We also see in Figure~\ref{fig:tissue} that the acid-induced tissue degradation is qualitatively similar in the two experiments, although quantitatively slightly lower in this experiment.}\\[-2ex]
	
	{\cb In contrast to Experiment}~\ref{exp:dom-hap} and the corresponding simulations in Figure~\ref{fig:exp1},  in the current experiment the glioma cell density drops significantly at the initial tumor location, while at the same time spreads further away from it. Since the diffusion-related weight parameter $\rho_2$ is the same between the two experiments, as do the rest of the parameters and modeling assumptions, {\cb this suggests that the observed difference in glioma invasion is due to the repellent pH-taxis and its domination over haptotaxis}. 
\end{experiment}

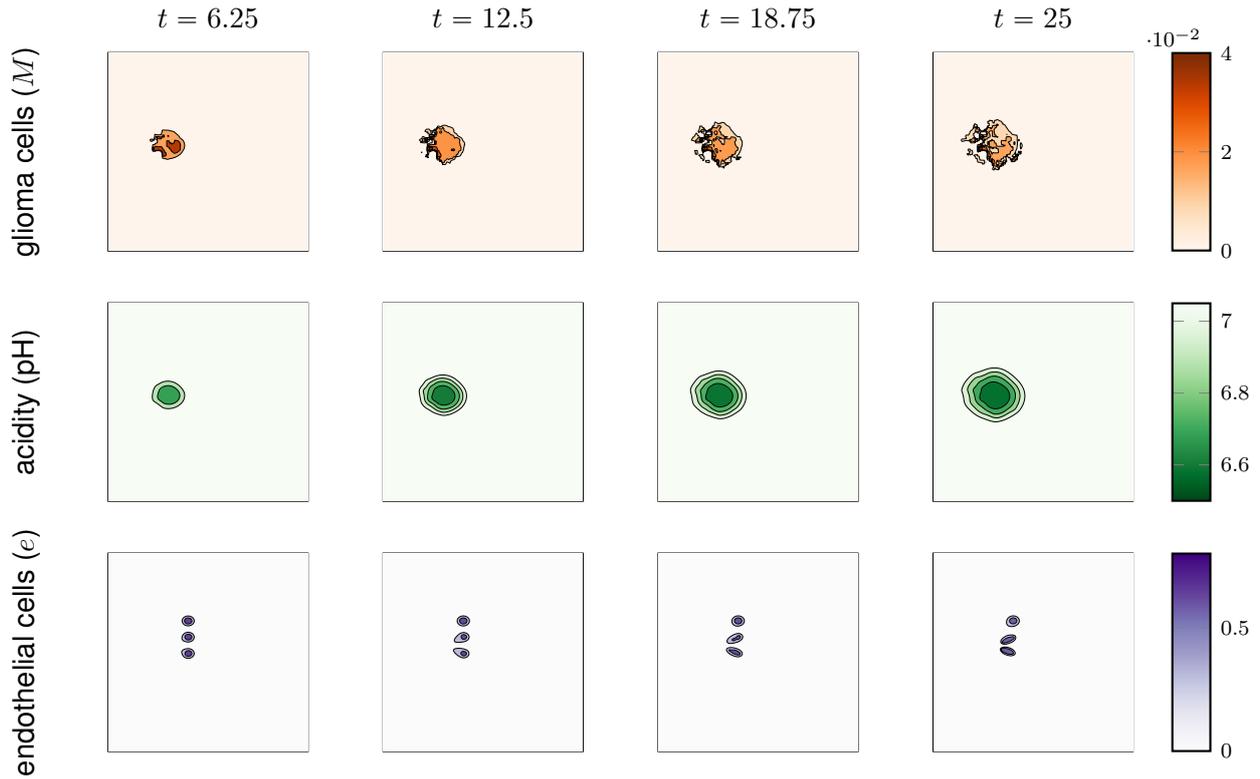
\begin{figure}[t]
	\figurewidth=\linewidth
	\def \expSetup {reference_CNN_unlimited}
	\begin{tikzpicture}
		\begin{groupplot}[
			/tikz/mark size=1.5pt,
			group style={
				group name=my plots,
				group size=4 by 3,
				horizontal sep=1cm,      
				vertical sep=0.7cm,        
			},
			xmin=-2,
			xmax=2,
			ymin=-2,
			ymax=2,
			ylabel shift = 2 em,
			xtick = \empty,
			ytick= \empty,
			ticklabel style = {font=\scriptsize},
			axis line style = thick,
			axis background/.style={fill=white},
			width=.25\figurewidth,
			height=.25\figurewidth, 
			]
			\nextgroupplot[ylabel = glioma cells ($M$), title = {$t=6.25$}]
			\addplot [forget plot] graphics [xmin=-2, xmax=2, ymin=-2, ymax=2] {\expSetup1};
			\nextgroupplot[title = {$t=12.5$}]
			\addplot [forget plot] graphics [xmin=-2, xmax=2, ymin=-2, ymax=2] {\expSetup2};
			\nextgroupplot[title = {$t=18.75$}]
			\addplot [forget plot] graphics [xmin=-2, xmax=2, ymin=-2, ymax=2] {\expSetup3};
			\nextgroupplot[
			title = {$t=25$},
			point meta min=0,
			point meta max=0.04,
			colormap name=glioma,
			colorbar,
			colorbar style={at={(1.2,1)},anchor=north west}
			]
			\addplot [forget plot] graphics [xmin=-2, xmax=2, ymin=-2, ymax=2] {\expSetup4};
			\nextgroupplot[ylabel = acidity (pH)]
			\addplot [forget plot] graphics [xmin=-2, xmax=2, ymin=-2, ymax=2] {\expSetup9};
			\nextgroupplot
			\addplot [forget plot] graphics [xmin=-2, xmax=2, ymin=-2, ymax=2] {\expSetup10};
			\nextgroupplot
			\addplot [forget plot] graphics [xmin=-2, xmax=2, ymin=-2, ymax=2] {\expSetup11};
			\nextgroupplot[
			point meta min=6.5,
			point meta max=7.05,
			colormap name=AC,
			colorbar,
			colorbar style={at={(1.2,1)},anchor=north west}
			]
			\addplot [forget plot] graphics [xmin=-2, xmax=2, ymin=-2, ymax=2] {\expSetup12};
			\nextgroupplot[ylabel = endothelial cells ($e$)]
			\addplot [forget plot] graphics [xmin=-2, xmax=2, ymin=-2, ymax=2] {\expSetup13};
			\nextgroupplot
			\addplot [forget plot] graphics [xmin=-2, xmax=2, ymin=-2, ymax=2] {\expSetup14};
			\nextgroupplot
			\addplot [forget plot] graphics [xmin=-2, xmax=2, ymin=-2, ymax=2] {\expSetup15};
			\nextgroupplot[
			point meta min=0,
			point meta max=0.8,
			colormap name=endothelial,
			colorbar,
			colorbar style={at={(1.2,1)},anchor=north west}
			]
			\addplot [forget plot] graphics [xmin=-2, xmax=2, ymin=-2, ymax=2] {\expSetup16};
		\end{groupplot}
	\end{tikzpicture}%
	\caption{Simulation results {\cb for} \textbf{Experiment~\ref{exp:unlimited} --- no flux limitation}. A major {\cb modification} of model~\eqref{full-system-macro-nondim} and Experiment~\ref{exp:dom-hap} by replacing the {\cb saturated} flux ~\eqref{coeff-M-eq-final-b-part} with {\cb the 'classical'} version~\ref{eq:non-limit-adapt}. {\cb The effect} is a much higher level of spatial tumor fragmentation {\cb (in particular exhibiting more irregular margins)} and {\cb less spread} than in Figure~\ref{fig:exp1}.}\label{fig:exp3}
\end{figure}

\begin{experiment}\textbf{no flux limitation.}\label{exp:unlimited}
	In this experiment {\cb we modify model~\eqref{full-system-macro-nondim} by removing} the flux limitation; this is obtained by replacing~\eqref{coeff-M-eq-final-b-part} with
	\begin{equation}\label{eq:non-limit-adapt}
		b(y^*)=-(1-\rho_1-\rho_2) \nabla h +\rho_1(1-y^*) \nabla Q-\rho_2 \nabla M.
	\end{equation}
	{\cb The rest of} the model components, initial conditions, and parameters are as set in Experiment~\ref{exp:dom-hap} and Table~\ref{tbl:ref_pars}. The corresponding simulation results are shown in Figure~\ref{fig:exp3}. When compared with Experiment~\ref{exp:dom-hap} and Figure~\ref{fig:exp1}, they reveal a qualitatively similar evolution of the acidity and a  similar vascularization pattern. The same holds true when comparing the degradation of the brain tissue between the two experiments; this is seen in Figure~\ref{fig:tissue}. The tumor, however, exhibits in the current experiment a clearly higher spatial fragmentation, {\cb with more fractal margins (which are characteristic for glioblastoma, see e.g. \cite{Gerstner}) and a more confined invasion. This is actually the expected effect of flux-saturated motility which eludes, among others, the nonphysical infinite speed of propagation typically connected with linear diffusion.} 
\end{experiment}

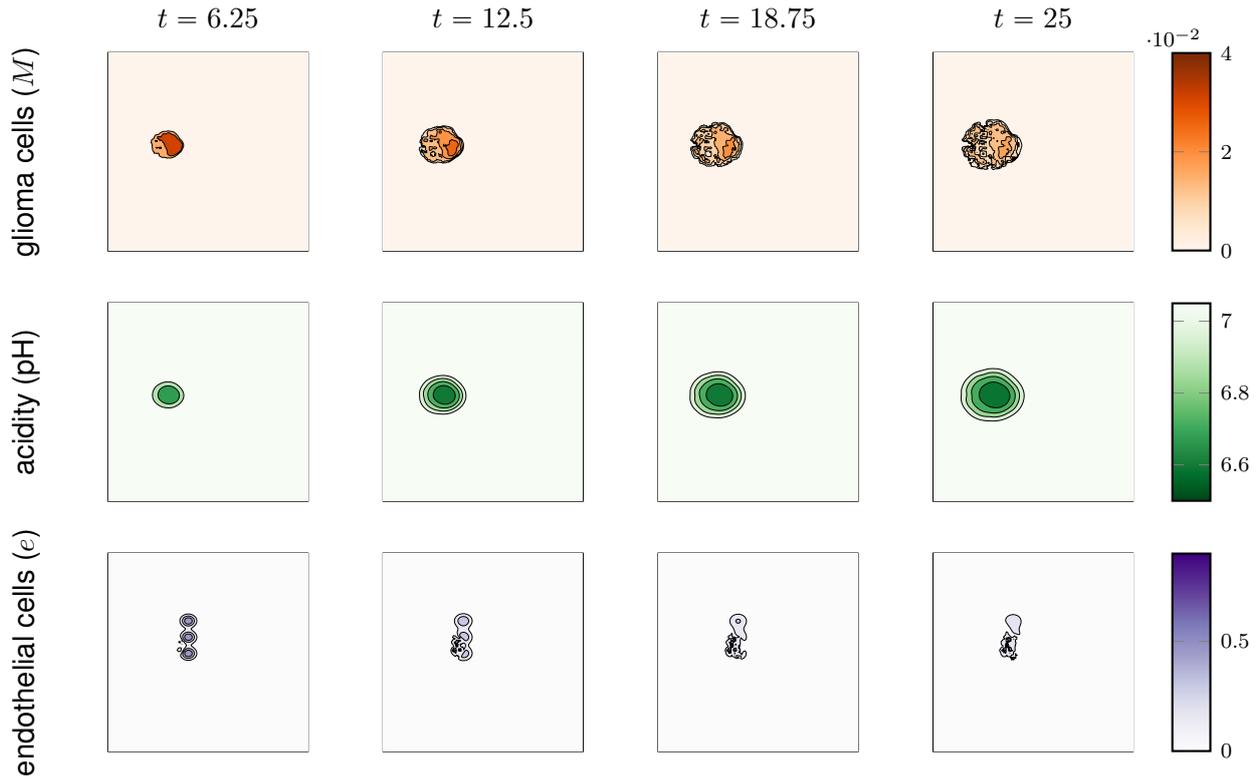
\begin{figure}[t] 
	\figurewidth=\linewidth
	\def \expSetup {reference_CNN_ge_attract}
	\begin{tikzpicture}
		\begin{groupplot}[
			/tikz/mark size=1.5pt,
			group style={
				group name=my plots,
				group size=4 by 3,
				horizontal sep=1cm,      
				vertical sep=0.7cm,        
			},
			xmin=-2,
			xmax=2,
			ymin=-2,
			ymax=2,
			ylabel shift = 2 em,
			xtick = \empty,
			ytick= \empty,
			ticklabel style = {font=\scriptsize},
			axis line style = thick,
			axis background/.style={fill=white},
			width=.25\figurewidth,
			height=.25\figurewidth, 
			]
			\nextgroupplot[ylabel = glioma cells ($M$), title = {$t=6.25$}]
			\addplot [forget plot] graphics [xmin=-2, xmax=2, ymin=-2, ymax=2] {\expSetup1};
			\nextgroupplot[title = {$t=12.5$}]
			\addplot [forget plot] graphics [xmin=-2, xmax=2, ymin=-2, ymax=2] {\expSetup2};
			\nextgroupplot[title = {$t=18.75$}]
			\addplot [forget plot] graphics [xmin=-2, xmax=2, ymin=-2, ymax=2] {\expSetup3};
			\nextgroupplot[
			title = {$t=25$},
			point meta min=0,
			point meta max=0.04,
			colormap name=glioma,
			colorbar,
			colorbar style={at={(1.2,1)},anchor=north west}
			]
			\addplot [forget plot] graphics [xmin=-2, xmax=2, ymin=-2, ymax=2] {\expSetup4};
			\nextgroupplot[ylabel = acidity (pH)]
			\addplot [forget plot] graphics [xmin=-2, xmax=2, ymin=-2, ymax=2] {\expSetup9};
			\nextgroupplot
			\addplot [forget plot] graphics [xmin=-2, xmax=2, ymin=-2, ymax=2] {\expSetup10};
			\nextgroupplot
			\addplot [forget plot] graphics [xmin=-2, xmax=2, ymin=-2, ymax=2] {\expSetup11};
			\nextgroupplot[
			point meta min=6.5,
			point meta max=7.05,
			colormap name=AC,
			colorbar,
			colorbar style={at={(1.2,1)},anchor=north west}
			]
			\addplot [forget plot] graphics [xmin=-2, xmax=2, ymin=-2, ymax=2] {\expSetup12};
			\nextgroupplot[ylabel = endothelial cells ($e$)]
			\addplot [forget plot] graphics [xmin=-2, xmax=2, ymin=-2, ymax=2] {\expSetup13};
			\nextgroupplot
			\addplot [forget plot] graphics [xmin=-2, xmax=2, ymin=-2, ymax=2] {\expSetup14};
			\nextgroupplot
			\addplot [forget plot] graphics [xmin=-2, xmax=2, ymin=-2, ymax=2] {\expSetup15};
			\nextgroupplot[
			point meta min=0,
			point meta max=0.9,
			colormap name=endothelial,
			colorbar,
			colorbar style={at={(1.2,1)},anchor=north west}
			]
			\addplot [forget plot] graphics [xmin=-2, xmax=2, ymin=-2, ymax=2] {\expSetup16};
		\end{groupplot}
	\end{tikzpicture}%
	\caption{Simulation results {\cb for} \textbf{Experiment~\ref{exp:ge-attract} --- unilateral interspecies attraction}. A major {\cb modification} of model~\eqref{full-system-macro-nondim} and Experiment~\ref{exp:dom-hap} by {\cb letting the endothelial cells follow gradients of glioma density instead of acidity gradients. Hence,}~\eqref{full-system-macro-nondim-e} was replaced with~\eqref{eq:adapt-endoth}. The main effect is on the vascularization, {\cb which is less directed and less pronounced} than in Figure~\ref{fig:exp1}.}\label{fig:exp4}
\end{figure}

\begin{experiment}\textbf{unilateral interspecies attraction.}\label{exp:ge-attract}
	In this experiment we {\cb replace the indirect chemotaxis of endothelial cells towards acidity produced by the tumor with a direct attraction of the endothelial cells towards the neoplasm, i.e. let them follow gradients of glioma density.\footnote{In \cite{kolbe2020modeling} we proposed another model for tumor invasion with multiple taxis and unilateral interspecies repellence, considering the go-or-grow dichotomy (also encountered in glioma development) and letting the migrating cells move away from the proliferating phenotype.} Concretely, we replace}~\eqref{full-system-macro-nondim-e} with
	\begin{equation}\label{eq:adapt-endoth}
		\partial_te=D_e\Delta e-\varsigma_e\nabla\cdot\left(e(1-e)\nabla M\right)+G_e(h,M)e(1-e).
	\end{equation}
	To account for the fact {\cb that glioma cells are less diffusive than the protons they produce, we enhance the diffusion and decrease the tactic sensitivity} of the endothelial cells. Accordingly, we adjust the corresponding parameters to $D_e = 2 \times 10^{-5}$ and $\varsigma_e = 3 \times 10^{-2}$. The other parameters and initial conditions are as in Experiment~\ref{exp:dom-hap} and Table~\ref{tbl:ref_pars}. The corresponding simulation results are shown in Figure~\ref{fig:exp4} and Figure~\ref{fig:tissue} and exhibit glioma growth, acidity evolution, and brain tissue degradation that {\cb are} qualitative similar to Experiment~\ref{exp:dom-hap} and Figure~\ref{fig:exp1}, {\cb with a tumor core inferring less cell depletion and the tumor mass showing a more homogeneous structure than that in Figure  ~\ref{fig:exp2} and lower cell densities than that in Figure~\ref{fig:exp1}.} On the other hand, the vascularization {\cb is in this case less directed and less pronounced than} in Experiment~\ref{exp:dom-hap}. {\cb Instead, the endothelial cells seem to leave their original sites and migrate in a rather diffusion-dominated way, occasionally forming smaller aggregates of high density.} 
\end{experiment}

\begin{experiment}\textbf{dominant haptotaxis in 3D}\label{exp:3d}
	The final numerical experiment considers the full brain in 3D and is conducted over the cuboid domain $\Omega_2=[0,1]\times[0,1.2155]\times[1.069]$ and the time interval $t\in[0,150]$. The initial conditions for the glioma cells and the acidity are given, for every $(x, y, z)\in\Omega_2$, through
	\begin{subequations}\label{eq:ICs}
		\begin{align}
			M_0(x,y,z) &=  0.1 e^{-\frac1\varepsilon \left( (x-0.63)^2+(y-0.608)^2 +(z-0.631)^2\right) },\\
			h_0(x,y) &= 0.03M_0(x,y) + 10^{-2.8},
		\end{align}
	\end{subequations}
	where $\varepsilon=5\times 10^{-5}$. The initial condition $Q_0$ {\cb for the spatial distribution of brain tissue density} is given in~\eqref{tissue_ini}. To account for the initial density of endothelial cells {\cb (which is not explicitly available in a typical DTI brain data set)} we have used {\cb -for simple illustrative purposes-} the random variables $e_0|_C= (10^{-2} + u_C)Q_0|_C$ on each numerical control volume $C\subset\Omega_2$, where $u_C\sim U(0,10^{-3})$ denote uniformly independent and identically distributed random variables for all control volumes $C$.  The parameters employed here are the same as in \textbf{Experiment~\ref{exp:dom-hap} --- dominant haptotaxis} and can be found in Table~\ref{tbl:ref_pars}.
	The simulation results are presented in Figure~\ref{fig:3d}. To allow for a better inspection of the glioma $M$ we have visualized it through its isosurface corresponding to a density of $10^{-5}$. Panel (a) shows the time evolution of the glioma $M$ in the brain tissue $Q$, from which a part has been extracted for the illustration. A closer inspection in panels (d) and (e) reveals the non-uniform growth and the dynamical adaptation of the glioma $M$ as seen also in the 2D Experiments {\cb \ref{exp:dom-hap}--\ref{exp:ge-attract}}. Protrusion of the glioma to the surrounding tissue is evident. Also, the pH level in the neighbourhood of the glioma is decreased similarly to the 2D experiments as shown in panel (c).  Contrary to the previous experiments, the {\cb density $e$ of} endothelial cells account{\cb s} for a full 3D vasculature of the brain in this experiment. Panel (b) shows the evolution of their density.

\begin{figure}
	\centering
	\begin{tabular}{cccc}
		\includegraphics[height=9em]{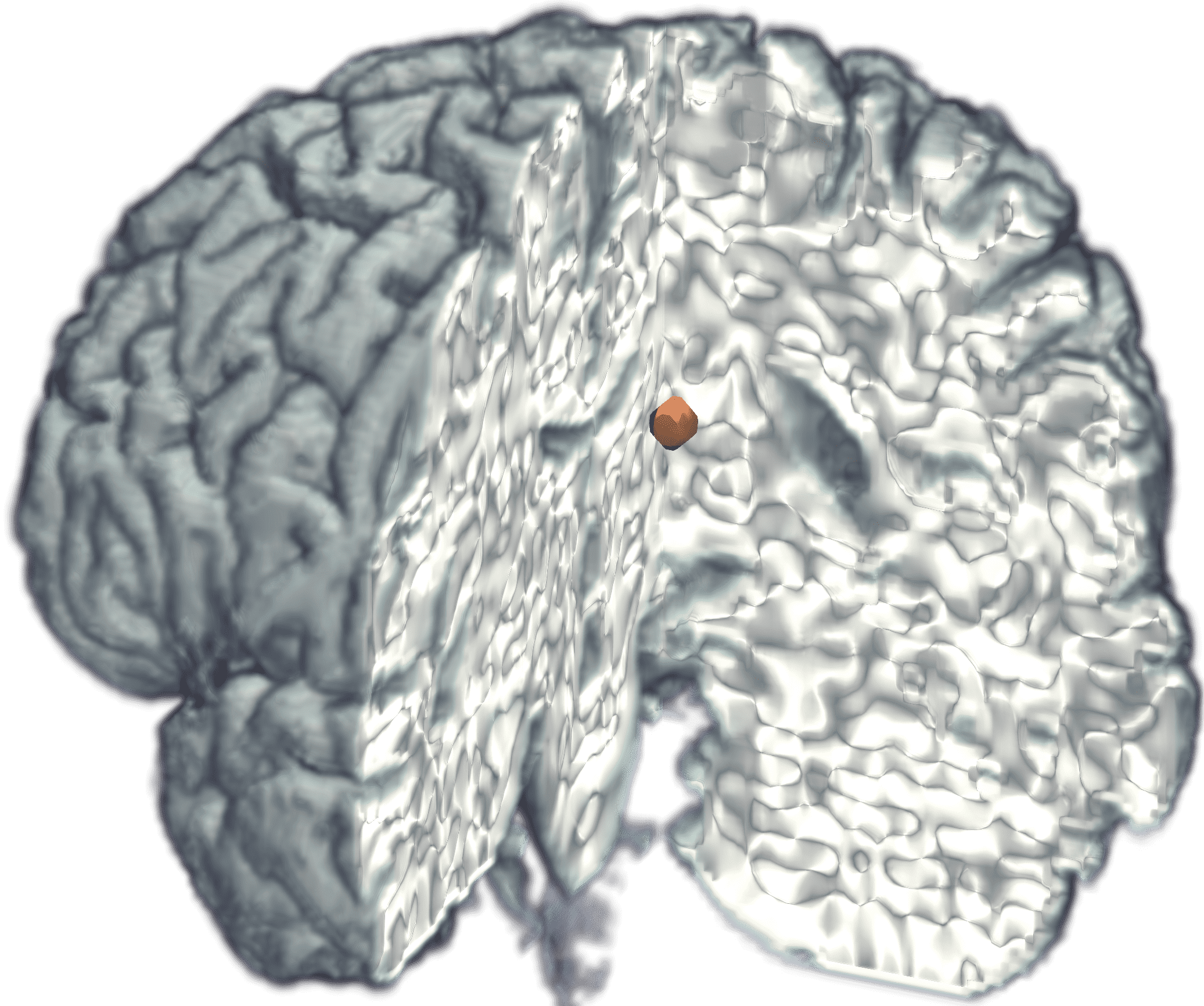}
			&\includegraphics[height=9em]{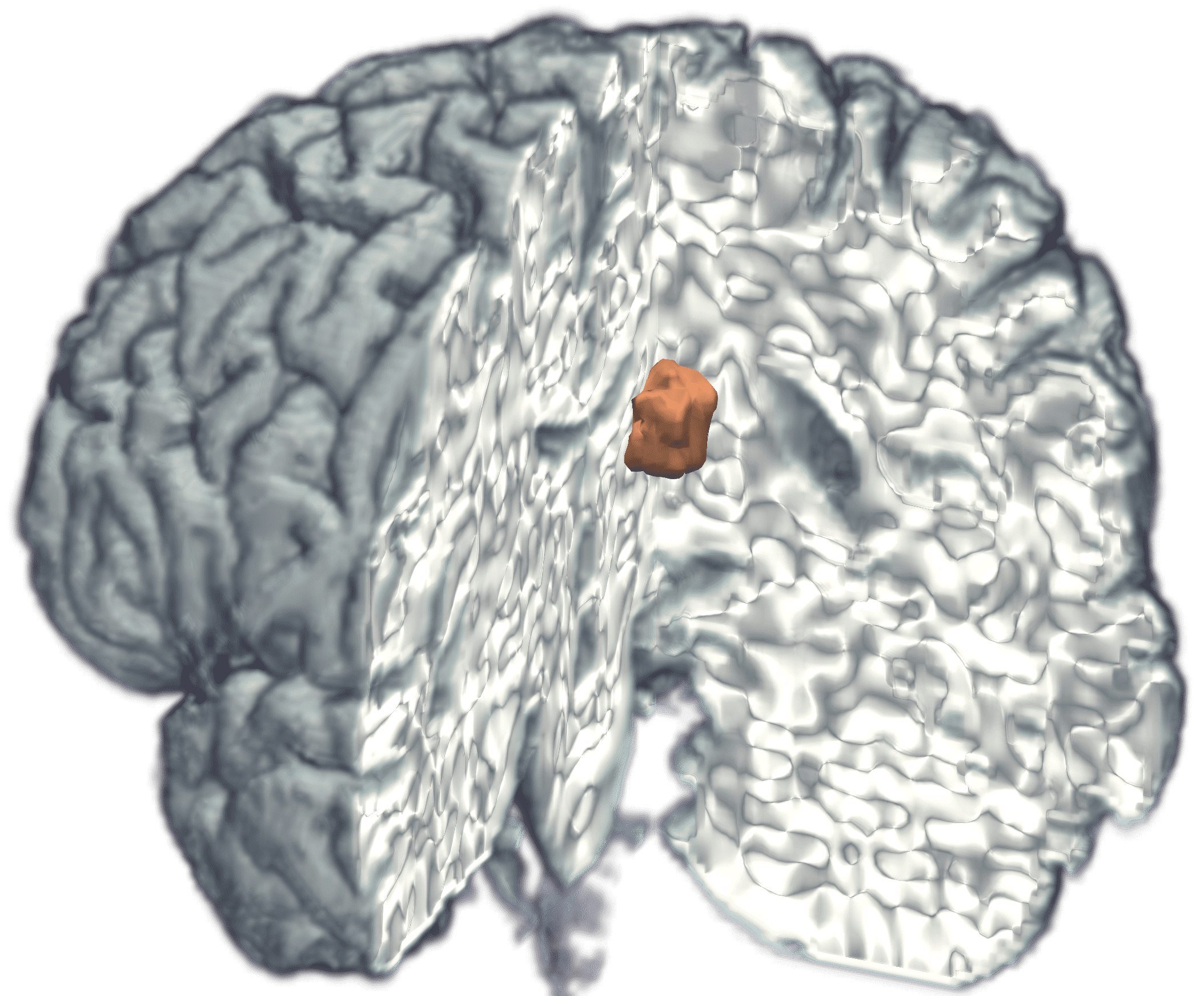}
			&\includegraphics[height=9em]{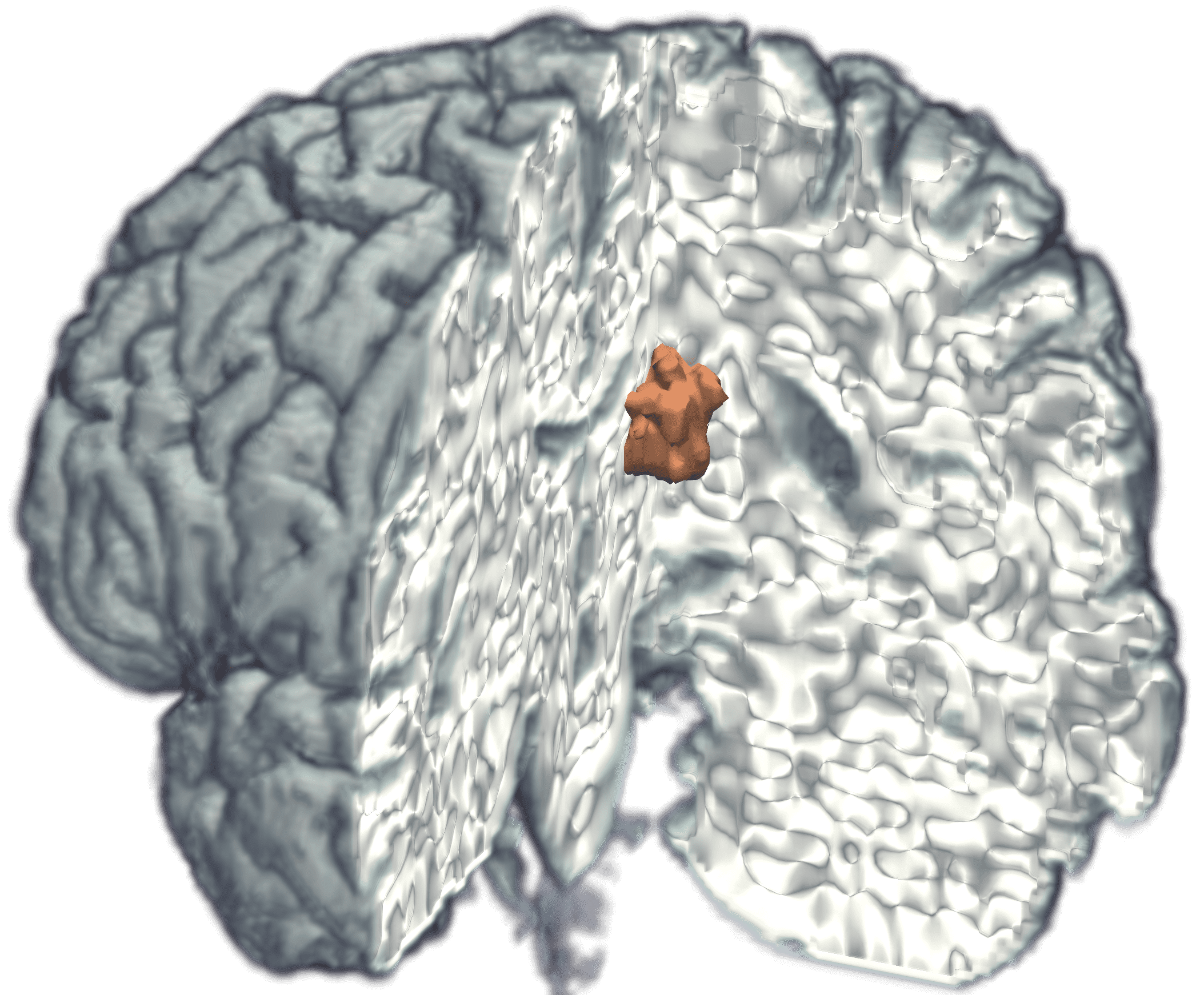}
			~\includegraphics[height=9em]{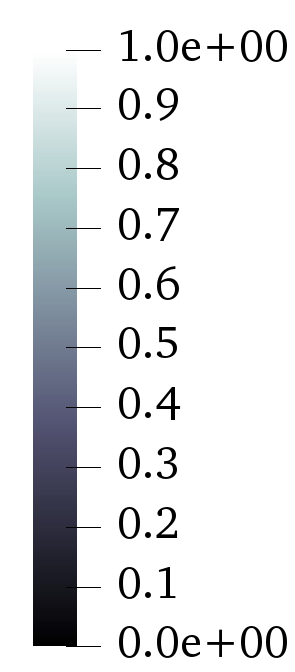}
		\\
		\multicolumn{3}{c}{(a) Tissue $Q$ and the $10^{-5}$ isosurface of the glioma $M$ at $t=0$, $t=75$, and $t=150$}	
		\\
		\includegraphics[height=10em]{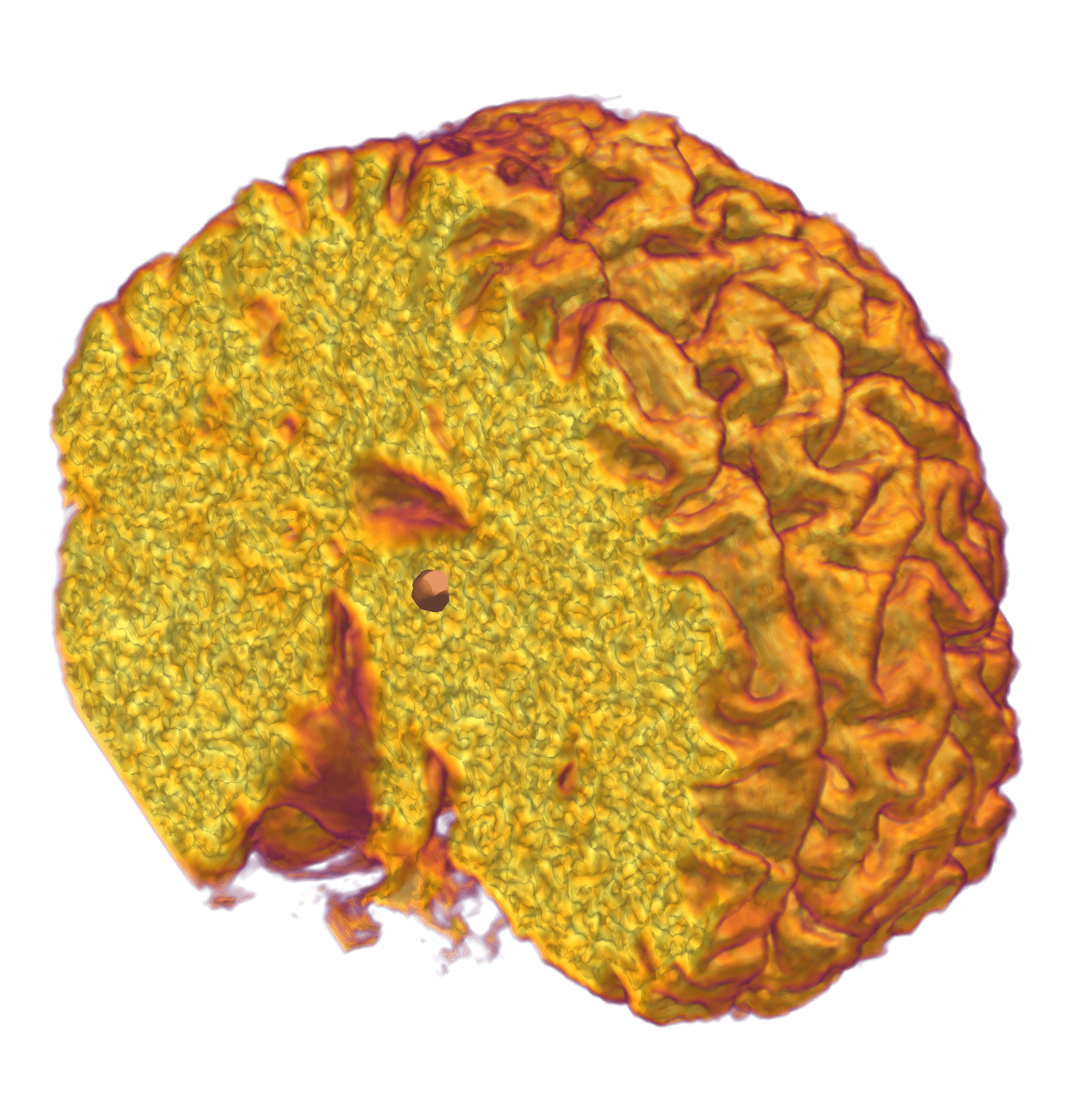}
			&\includegraphics[height=10em]{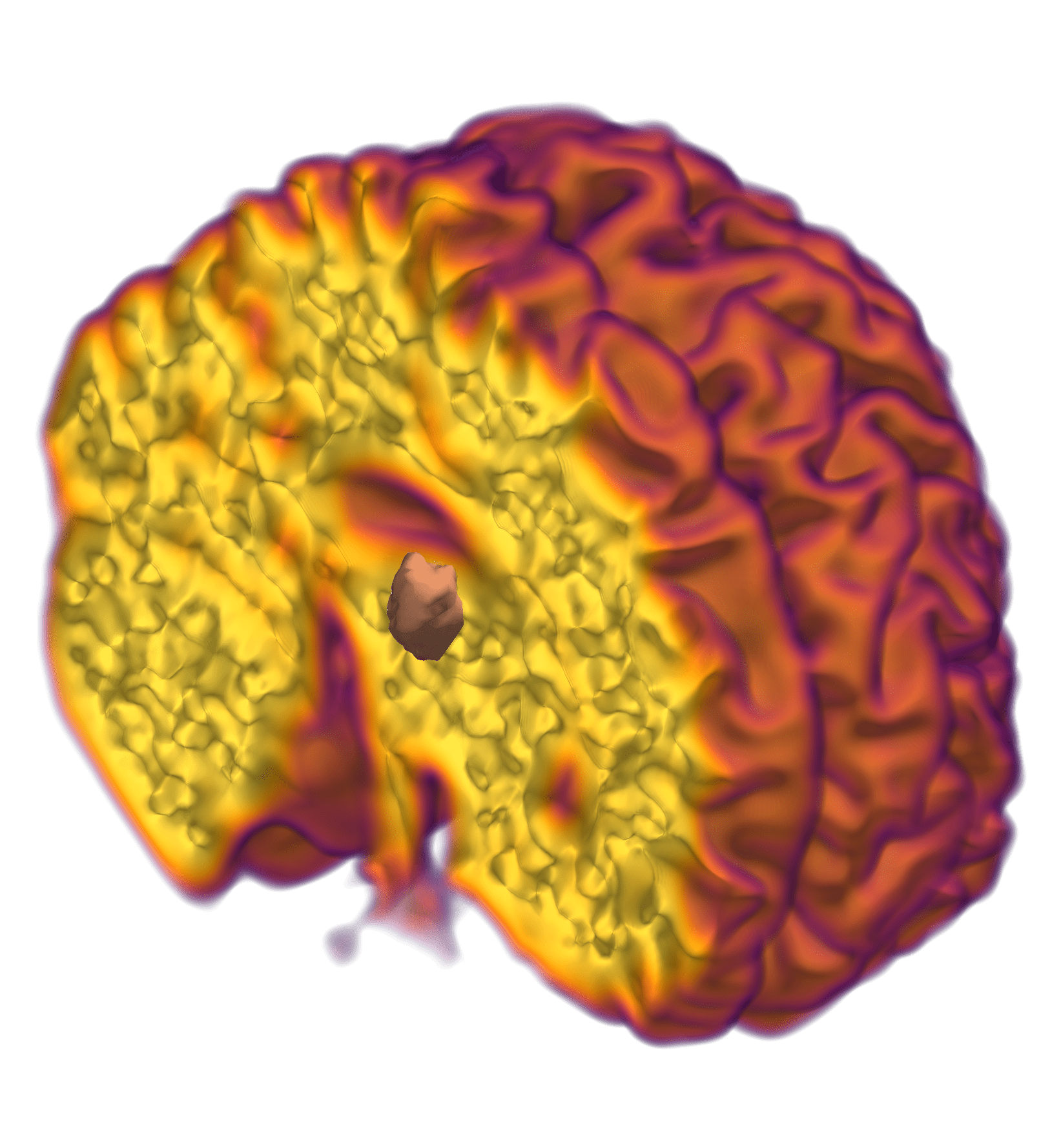}
			&\includegraphics[height=10em]{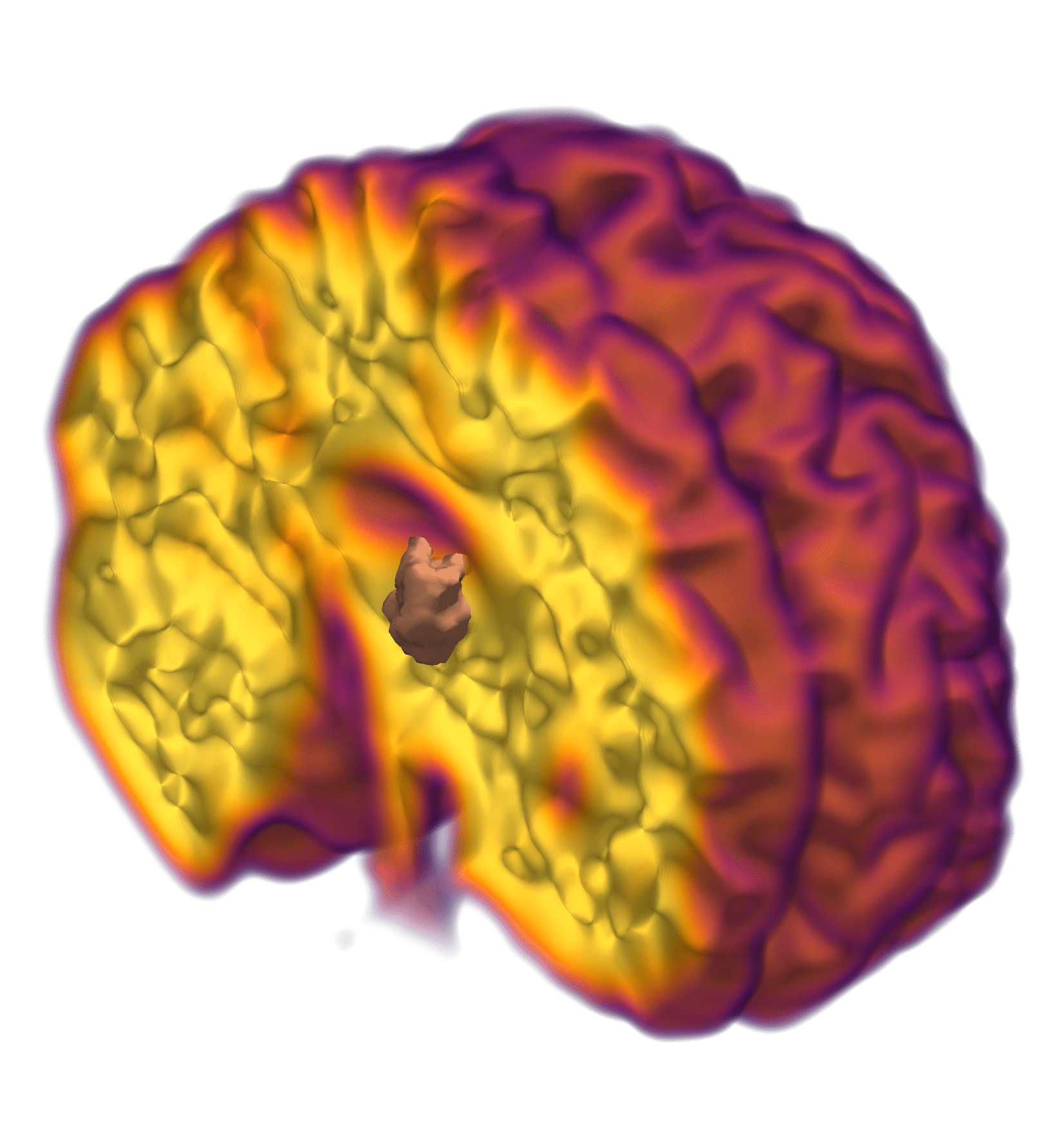}~\includegraphics[height=9em]{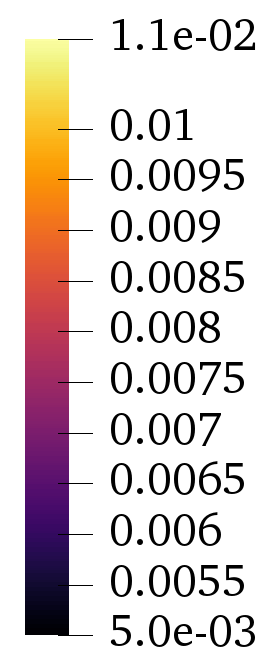}
		\\
		\multicolumn{3}{c}{(b) Vasculature $e$ and the $10^{-5}$ isosurface of the glioma $M$ at $t=0$, $t=75$, and $t=150$}
		\\
		\includegraphics[height=8em]{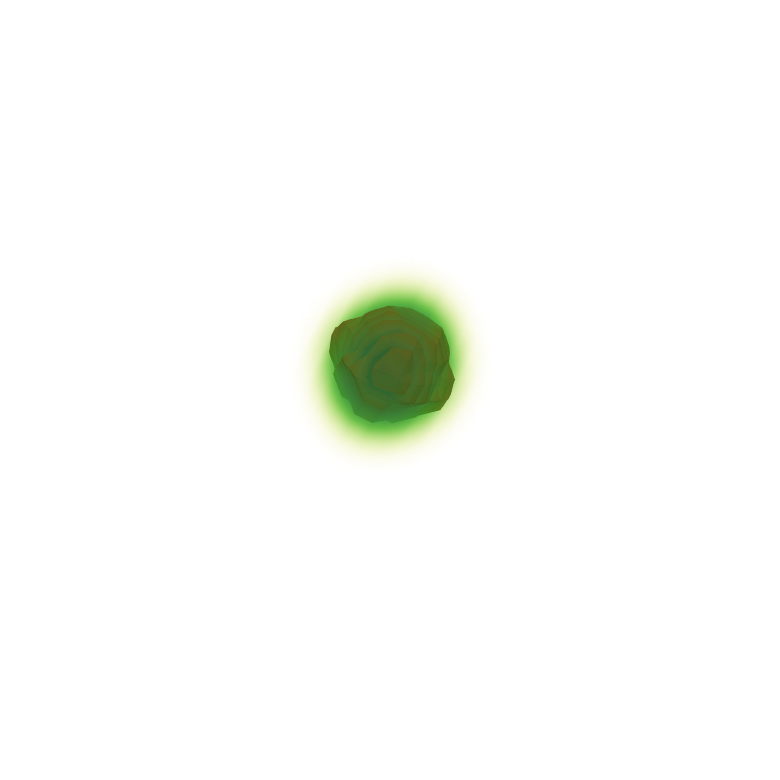}
			&\includegraphics[height=8em]{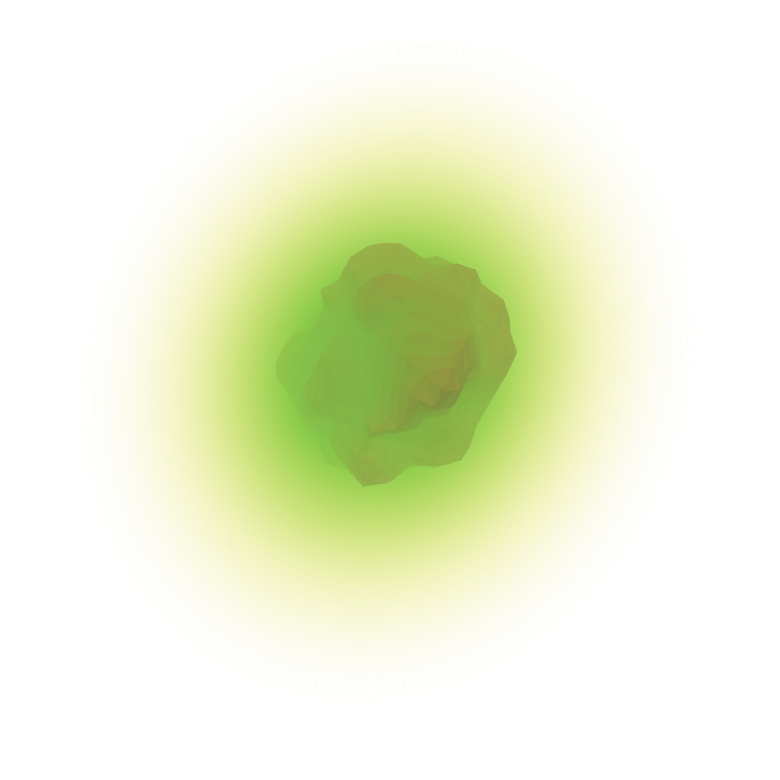}
			&\includegraphics[height=10em]{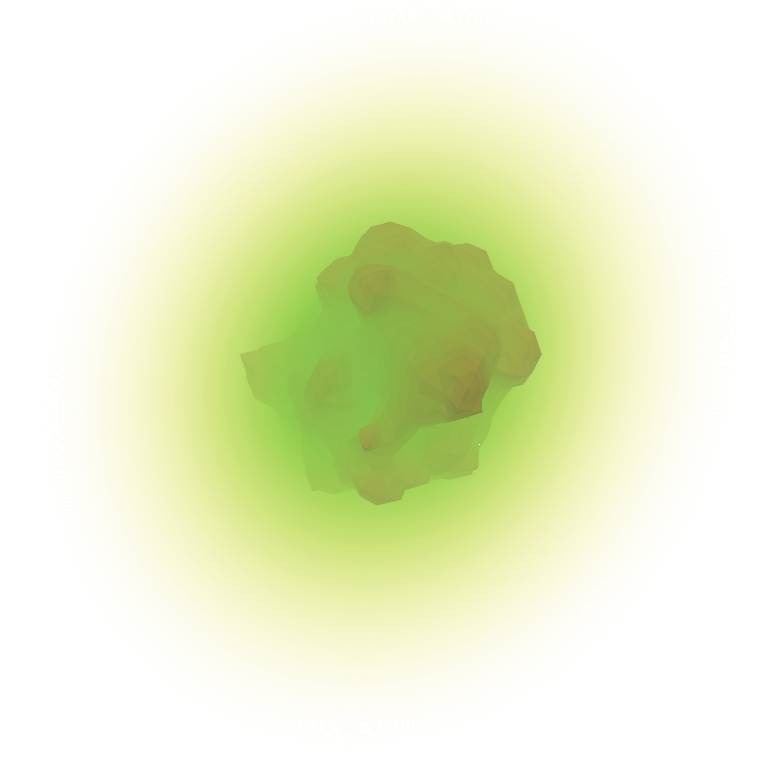}~\includegraphics[height=9em]{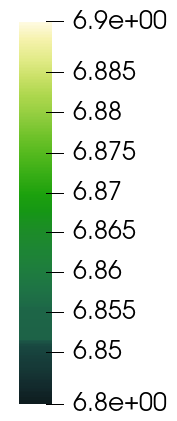}
		\\
		\multicolumn{3}{c}{(c) Acidity (pH) and the $10^{-5}$ isosurface of the glioma $M$ at $t=0$, $t=75$, and $t=150$}
		\\
		\multicolumn{2}{c}{
			\includegraphics[height=9em]{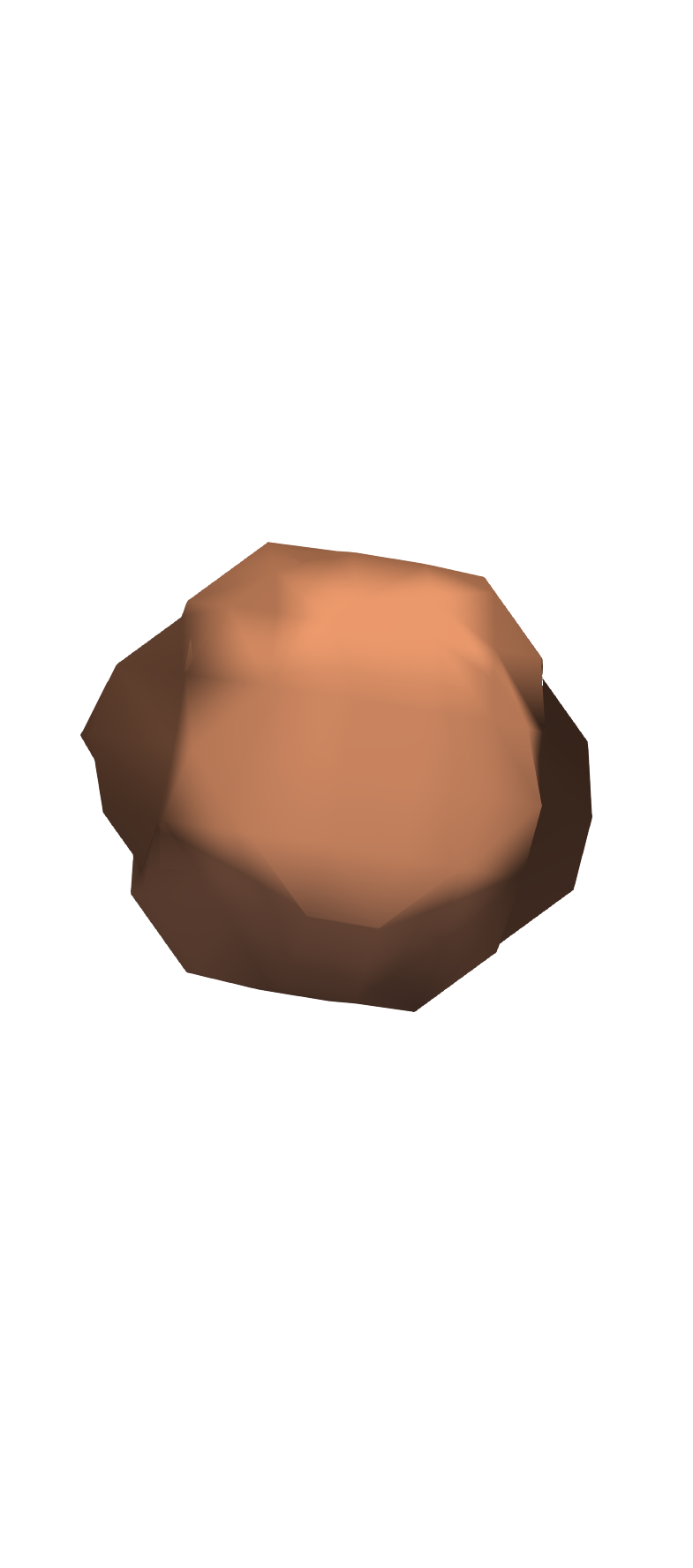}
			~\includegraphics[height=9em]{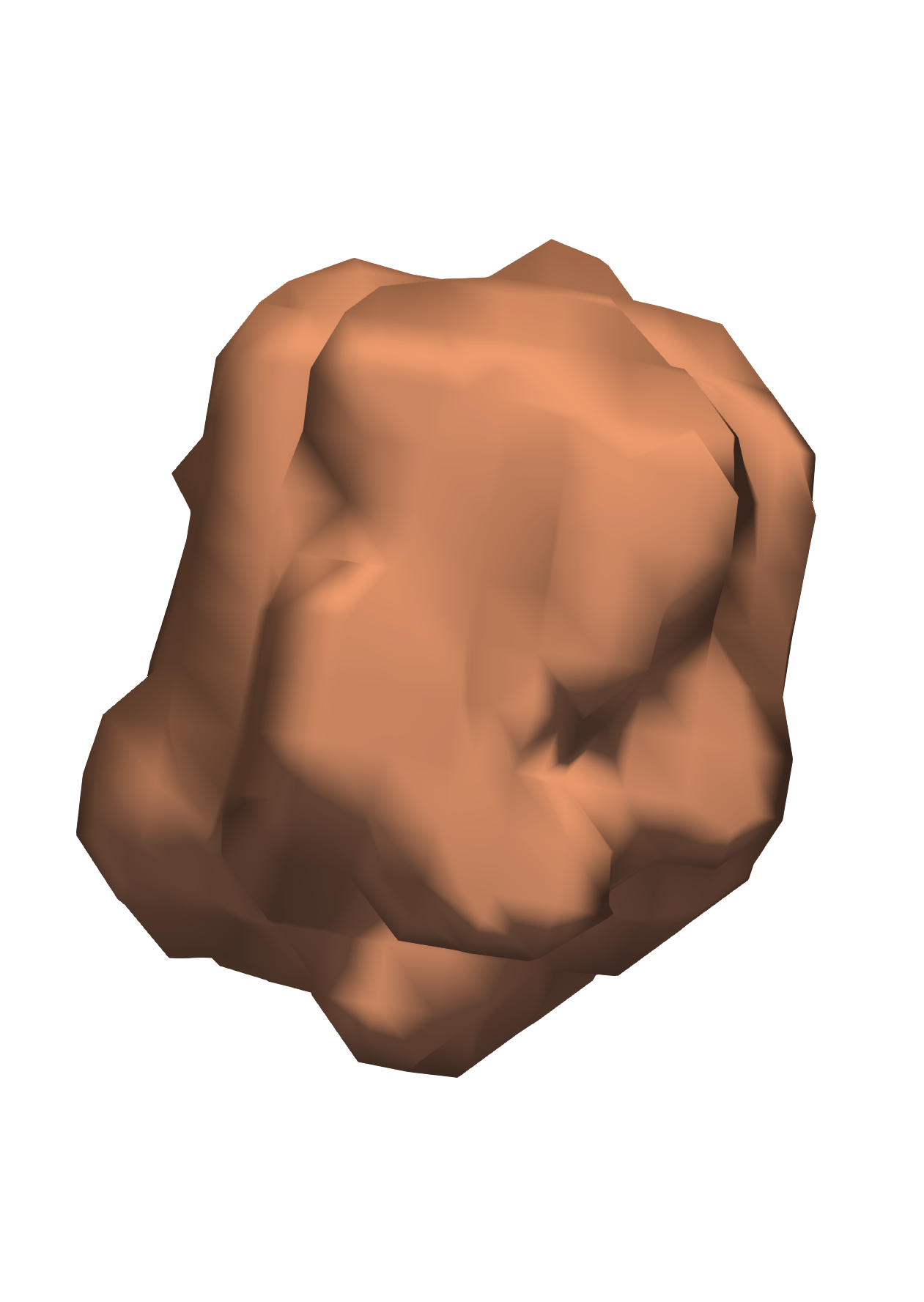}
			~\includegraphics[height=9em]{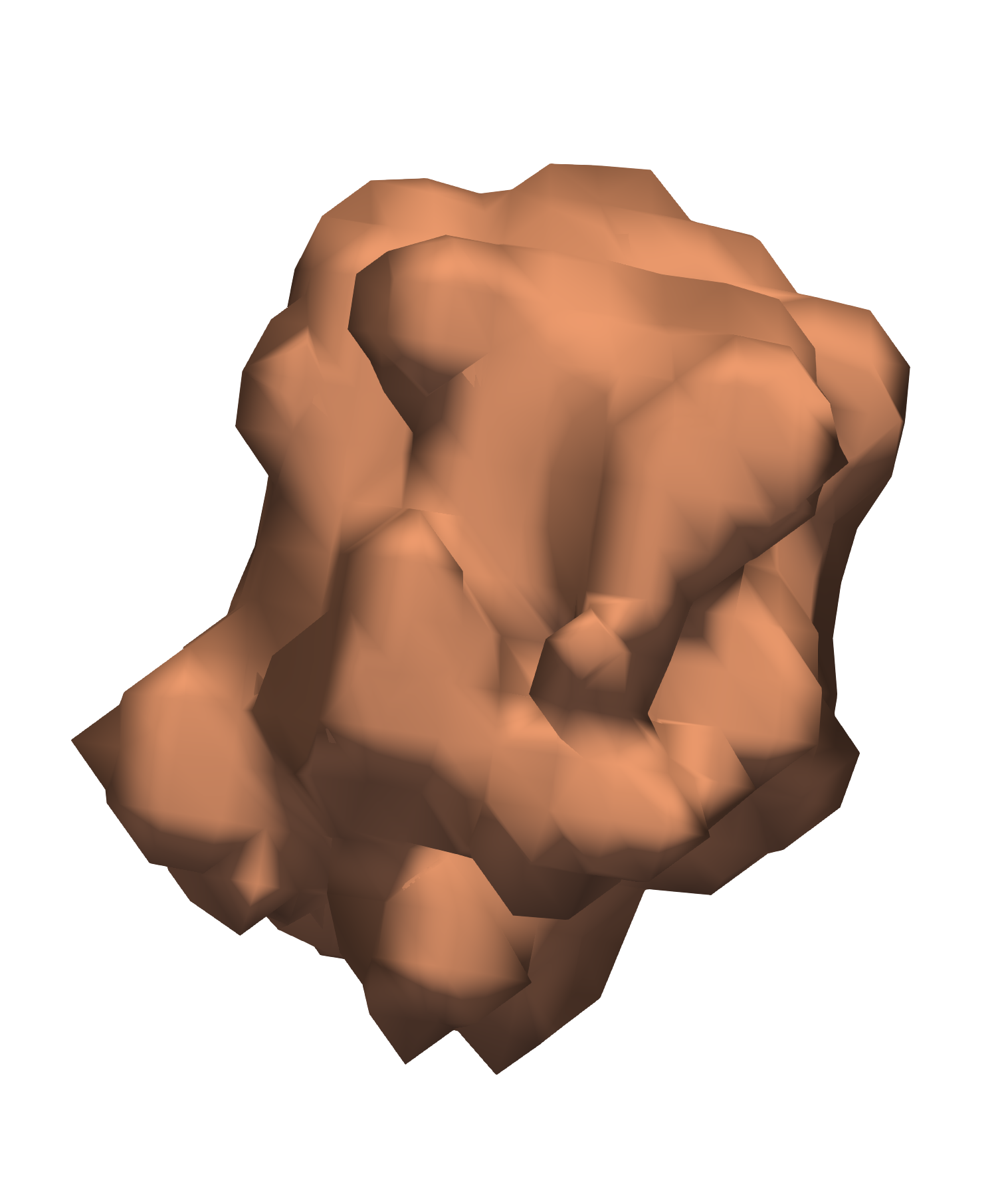}}
		&\includegraphics[height=8em]{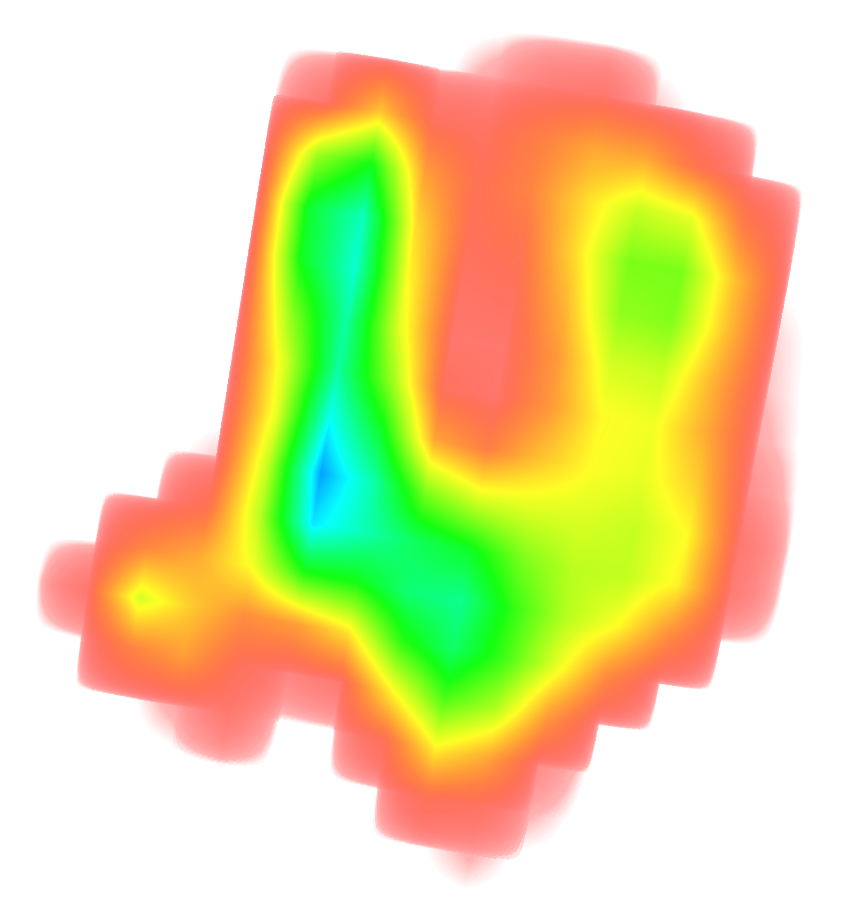}
	   ~\includegraphics[height=9em]{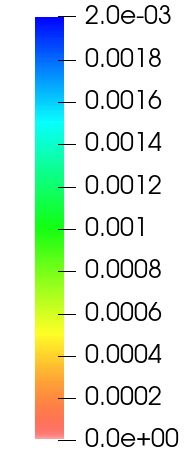}
		\\
		\multicolumn{2}{c}{(d) Glioma $M$ at $t=0$, $t=75$, $t=150$}&(e) Glioma $M$ inner structure at $t=150$
		\\
		
	\end{tabular}
\caption{Simulation results for \textbf{Experiment~\ref{exp:3d}}. (a): Time evolution of the brain tissue $Q$ along with the $10^{-5}$ isosurface of the glioma $M$. The tissue regeneration and degradation processes are in action although their effects are not very visible. {The colorbar on the right corresponds to the tissue $Q$.} The {tumor} grows significantly in size and attains the particular geometric conformation seen in (d).  (c): Time evolution of the vasculature $e$ along with the $10^{-5}$ isosurface {of the glioma} $M$. 
(c): The time evolution of the pH, along with the $10^{-5}$ isosurface of glioma $M$, shows the way the acid $h$ spreads through the tissue. (d): Close-up of the $10^{-5}$ isosurface of the  tumour at three different time instances. (e): A plane-cut through the tumour at $t=150$ reveals the regions and distribution of higher tumour densities.}\label{fig:3d}
\end{figure}
\end{experiment}

\mysection{Discussion}\label{sec:discussion}

The bottom-up modeling approach proposed here is inspired by \cite{CEKNSSW} and also related to the simplified earlier setting in \cite{Chauviere2007}, but differs from those formulations by the way in which the upscaling was performed and, essentially, by the form of the obtained macroscopic PDE for glioma density evolution, which features flux-limited self-diffusion, haptotaxis, and repellent pH-taxis. Moreover, the constant glioma cell speed assumption made in \cite{Chauviere2007,CEKNSSW} was relaxed, which influenced not only the scaling, but also the macroscopic motility and source terms. As mentioned in Section~\ref{sec:intro}, our approach leading to flux-saturated motility terms is different from that in \cite{BBNS2010,perthame2018flux}, since those terms originate here in the single-cell dynamics provided in~\eqref{vdynamics} and the corresponding transport term w.r.t. cell velocity in the KTE~\eqref{mesoscopic} rather than the cell turning operator. The method suggests that including (via Newton's second law) appropriate mechanical and chemical influences exerted on the cells can lead on the macroscale to yet other drift and/or diffusion terms, possibly with flux limitation. The deduction performed here is merely formal; a rigorous one, which follows a different limiting procedure and another form of flux saturation on the cell scale is addressed in a rigorous manner in \cite{ZS20}, where there is (tactic) flux limitation only in the macroscopic PDE for the first order correction. \\[-2ex]

The flux-saturated diffusion obtained in~\eqref{full-system-macro-nondim} eludes the nonphysical infinite speed of propagation and involves a nonlinearity accounting at least partially for intraspecific cell interactions. In contrast, the  model with flux-limited chemo- and haptotaxis formulated in \cite{kim2009} directly on the macroscale considers intrapopulation cell-cell interactions by way of an adhesion operator involving nonlocality w.r.t. space. In \cite{KPSZ} it has been recently proved that terms characterizing cell-cell and cell-tissue interactions described as spatial nonlocalities actually lead (in the rigorous limit of shrinking radius of the corresponding region) to taxis and self-diffusion. Other ways to model mutual cell interactions use avoidance of crowding in (some of) the motility and/or source terms, in a local or nonlocal manner, see \cite{CPSZ} for a review concerning settings with various types of nonlocalities. Lately, more attention has been attached to obtaining nonlocal kinetic models for cell migration characteristics depending on cell density \cite{loy2020stability}, some obtained, too, by   macroscopic limits \cite{loy2019kinetic,Loy2020}. In the present work the intrapopulation exchange is modeled on the one hand via logistic-type limitation of growth and on the other hand by accounting for changes in cell velocity orientation which are due to population pressure and motility limited by crowding. As such (besides flux saturations), our approach is yet different from \cite{loy2019kinetic,Loy2020,loy2020stability}, who do not account for single-cell velocity dynamics, but rather describe velocity and speed innovations by way of adequately chosen turning kernels and turning rates. \\[-2ex]

Systems with flux-limited diffusion and drift raise several challenges. Among others, the different structure of diffusion terms does not allow to directly apply the usual theoretical tools for handling parabolic PDEs, the solutions have poor regularity, possibly developing transient or even perpetual singularities; we refer to \cite{Calvo2015} for a review of (single) PDE models featuring flux limitations and their mathematical issues. Results about qualitative analysis of systems involving PDEs of reaction-diffusion-taxis type with one or several flux-saturated motility terms are unknown. Even  systems with multiple taxis of a more 'usual' kind (see \cite{kolbe2020modeling} for a very recent review) exhibit manifold challenges w.r.t. well-posedness and qualitative properties of their solutions, and we are not aware of any results concerning models of the type obtained in~\eqref{full-system-macro-nondim}, even if none of the terms in~\eqref{coeff-M-eq-final-b-part} would infer flux limitation.   



\addcontentsline{toc}{section}{References}
\bibliographystyle{plain}
\bibliography{literature}

\end{document}